\documentclass[12pt,a4paper]{article}

\usepackage{amsmath}
\usepackage{amsthm,amssymb}

\usepackage{graphicx} 

\newcommand{\Ref}[1]{(\ref{#1})}
\newcommand{\nn}{\nonumber \\}

\newcommand{\beq}{\begin{equation}}
\newcommand{\eeq}{\end{equation}}

\newcommand{\ds}{\displaystyle}

\newtheorem{thm}{Theorem}

\numberwithin{equation}{section}

\newcounter{mycase} 

\newenvironment{mycase}{\refstepcounter{mycase}}{}
\begin{document}
\title{Confining multiple polymers between sticky walls: a directed walk model of two polymers}
\author{Thomas Wong$^1$, Aleksander L Owczarek$^2$ and Andrew Rechnitzer$^1$ \\[1ex]
  \footnotesize
  \begin{minipage}{9cm}
  $^1$Department of Mathematics,\\
  University of British Columbia,\\
  Vancouver V6T 1Z2, British Columbia, Canada.\\
  \texttt{twong@math.ubc.ca,andrewr@math.ubc.ca}\\[1ex]
    $^2$Department of Mathematics and Statistics,\\
    The University of Melbourne, Victoria~3010, Australia.\\
    \texttt{owczarek@unimelb.edu.au}
\end{minipage}
}

\maketitle  

\begin{abstract}
We study a model of two polymers confined to a slit with sticky walls. More precisely, we find and analyse the exact solution of two directed friendly walks in such a geometry on the square lattice. We compare the \emph{infinite slit limit}, in which the length of the polymer (thermodynamic limit) is taken to infinity before the width of the slit is considered to become large, to the opposite situation where the order of the limits are swapped, known as the \emph{half-plane limit} when one polymer is modelled. In contrast with the single polymer system we find that the half-plane and infinite slit limits coincide. We understand this result in part due to the tethering of polymers on both walls of the slit.

We also analyse the entropic force exerted by the polymers on the walls of the slit. Again the results differ significantly from single polymer models. In a single polymer system both attractive and repulsive regimes were seen, whereas in our two walk model only repulsive forces are observed. We do, however, see that the range of the repulsive force is dependent on the parameter values. This variation can be explained by the adsorption of the walks on opposite walls of the slit.

\end{abstract}
\pagebreak

\section{Introduction}

The adsorption of polymers on a sticky wall, or walls, and more recently the
pulling, or stretching, of a polymer away from a wall has been the subject of
continued interest \cite{privman1988c-a,debell1993a-a,janse2000a-a,rosa2003a-a,
orlandini2004a-a,krawczyk2004a-:a,brak2005a-:a,mishra2005a-a,janse2005a-:a,
martin2007a-:a}. This has been in part due to the advent of experimental
techniques able to micro-manipulate single polymers
\cite{svoboda1994a-a, ashkin1997a-a, strick2001a-a} and the connection to
modelling DNA denaturation \cite{essevaz-roulet1997a-a,
lubensky2000a-a,lubensky2002a-a, orlandini2001a-a, marenduzzo2002a-a,
marenduzzo2003a-a,marenduzzo2009a-a}.

When a polymer in a dilute solution of good solvent, so that it is in
a swollen state \cite{gennes1979a-a}, is then attached to a wall at
one end the rest of the polymer drifts away due to entropic
repulsion. It otherwise acts as if it were a free polymer. If the wall has an
attractive contact potential so that it becomes sticky to the
monomers the polymer can be made to stay close to the
wall by a sufficiently strong potential or at low enough temperatures.
The second-order phase transition between these two states is the
\emph{adsorption} transition. The high temperature state is
\emph{desorbed} while the low temperature state is \emph{adsorbed}. This pure
adsorption transition has been well studied
\cite{privman1988c-a,debell1993a-a,hegger1994a-a,janse2000a-a,janse2004a-a} exactly, and numerically, and
has been demonstrated to be second-order.

The situation becomes more complex when a polymer is confined between two 
sticky walls. This situation has been studied by various directed and 
non-directed lattice walk models \cite{brak2005a-:a,janse2005a-:a, 
brak2007motzkin, martin2007a-:a, owczarek2008a-:a, alvarez2008self, 
guttmann2009effect}. Here the phase diagram of the model can depend on the mesoscopic size of the polymer relative to the width of the slab/slit and the strengths of the interactions on both walls.   
A motivation for studying  this type of system is related to modelling  the stabilisation of colloidal dispersions by adsorbed  polymers (steric stabilisation) and the destabilisation when the polymer can 
adsorb on surfaces of different colloidal particles (sensitised 
flocculation). A polymer confined between two parallel plates exerts a repulsive force on the confining plates because of the loss of configurational entropy unless the polymer is attracted to both walls when it can exert an effective attractive force at large distances.

\begin{figure}[ht!]
\begin{center}
\includegraphics[width=10cm]{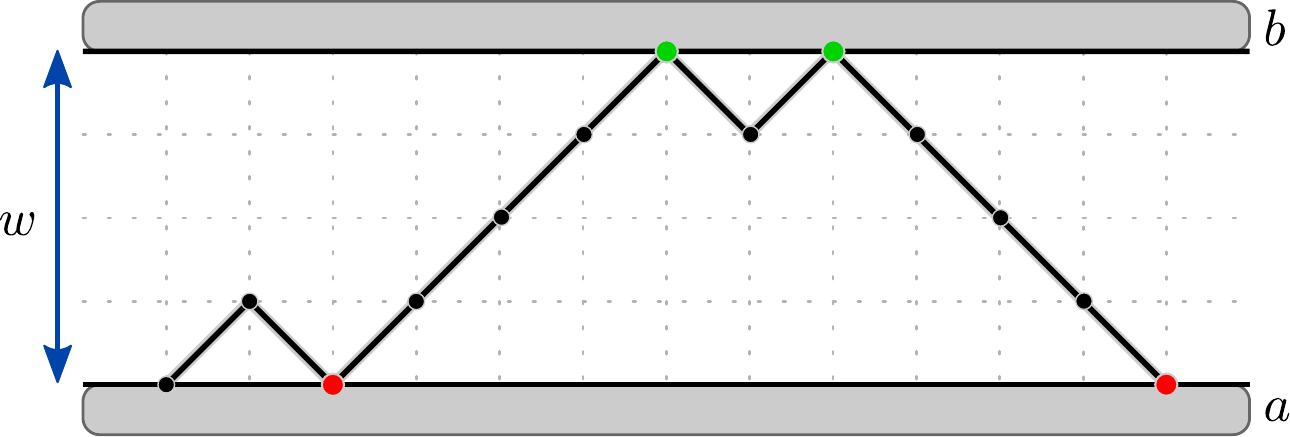}
\caption{ A Dyck path confined between two walls spaced $w$ lattice units 
apart. Each visit to the bottom wall contributes a Boltzmann weight $a$ 
and each visit to the top wall contributes a Boltzmann weight $b$. For 
combinatorial reasons we do not weight the first vertex.}
\label{onewalk} 
\end{center}
\end{figure}

A directed walk model of a polymer confined between two sticky walls was studied by Brak {\it et al.\ }\cite{brak2005a-:a}. Let us now briefly review the findings of that work so as to motivate the model we study in this paper. In their model the polymers are represented by Dyck paths, which are directed paths in the plane, taking north-east and south-east steps starting on, ending on and staying above the horizontal axis. These are classical objects in combinatorics \cite{flajolet2009analytic}. 
The height of these paths is then restricted; this is interpreted as  a model of  a polymer confined between two walls that are $w$ lattice units apart, as 
in Figure~\ref{onewalk}. It will be crucial to understand the results to note that the polymer is attached to the bottom wall at its end.  Finally,  
different Boltzmann weights $a$ and $b$ were added for each visit to the bottom 
and top walls respectively. 
The partition function for these paths is defined as
\begin{align}
Z_n^{single}(a,b;w) &=   \sum_{\varphi \in \mathcal{S}_w^n} a^{m_a(\varphi)} 
b^{m_b(\varphi)}\; ,
\end{align}
where $\mathcal{S}_w^n$ is the set of Dyck paths of length $n$ of restricted 
height with maximum $w$,  $m_a(\varphi)$ the number of vertices on the bottom 
wall and $m_b(\varphi)$ the number of vertices on the top wall (excluding the 
leftmost vertex).

A phase transition can only occur when both the thermodynamic limit  and the 
limit of infinite width (to give a two-dimensional thermodynamic system) are 
  taken. However, it was explained by Brak \emph{et al.} \cite{brak2005a-:a} 
that taking the thermodynamic limit $n\rightarrow\infty$ before or after taking 
the width of the slit to infinity is crucially important. If the width, $w$, of 
the system is taken to infinity first then the walk does not see the top wall 
and a half plane system is retrieved, since the polymer is tethered to the bottom wall. There is a simple adsorption transition as 
$a$ is varied: a second order phase transition occurs when $a=2$. On the other 
hand, if the thermodynamic limit is taken before the width is taken to infinity 
then a different phase diagram ensues dependent on both $a$ and $b$.   

We define 
the reduced free energy $\kappa^{single}(a,b;w)$ for single Dyck paths at fixed 
finite~$w$~as
\begin{align}
\kappa^{single}(a,b;w)  
&=  \lim_{n\rightarrow\infty} \frac{1}{n} \log Z_n^{single}(a,b;w)
\end{align}
and taking the limit $w\rightarrow \infty$ gives the so-called \emph{infinite 
slit} limit:
\begin{align}
\kappa^{single}_{inf-slit}(a,b) &
\equiv \lim_{w \rightarrow \infty} \kappa^{single}(a,b;w) = \lim_{w\rightarrow\infty} 
\lim_{n\rightarrow\infty}  \frac{1}{n}  \log Z_n^{single}(a,b;w)\,.
\label{inf-slit-free-energy-single-def}
\end{align}
This  limit is different from the \emph{half-plane limit} 
\begin{align}
 \kappa^{single}_{half-plane}(a) &= \lim_{n\rightarrow\infty} 
\lim_{w\rightarrow\infty}  \frac{1}{n} \log Z_n^{single}(a,b;w)= \begin{cases}
\log \left(2\right) & 
\mbox{ if } a
        \leq 2 \\ 
\log\left(\frac{a}{\sqrt{a-1}}\right) & \mbox{ if } a > 2
\end{cases}\, ,
 \end{align}
which is independent of $b$.

\begin{figure}[ht!]
\begin{center}
\includegraphics[width=8cm]{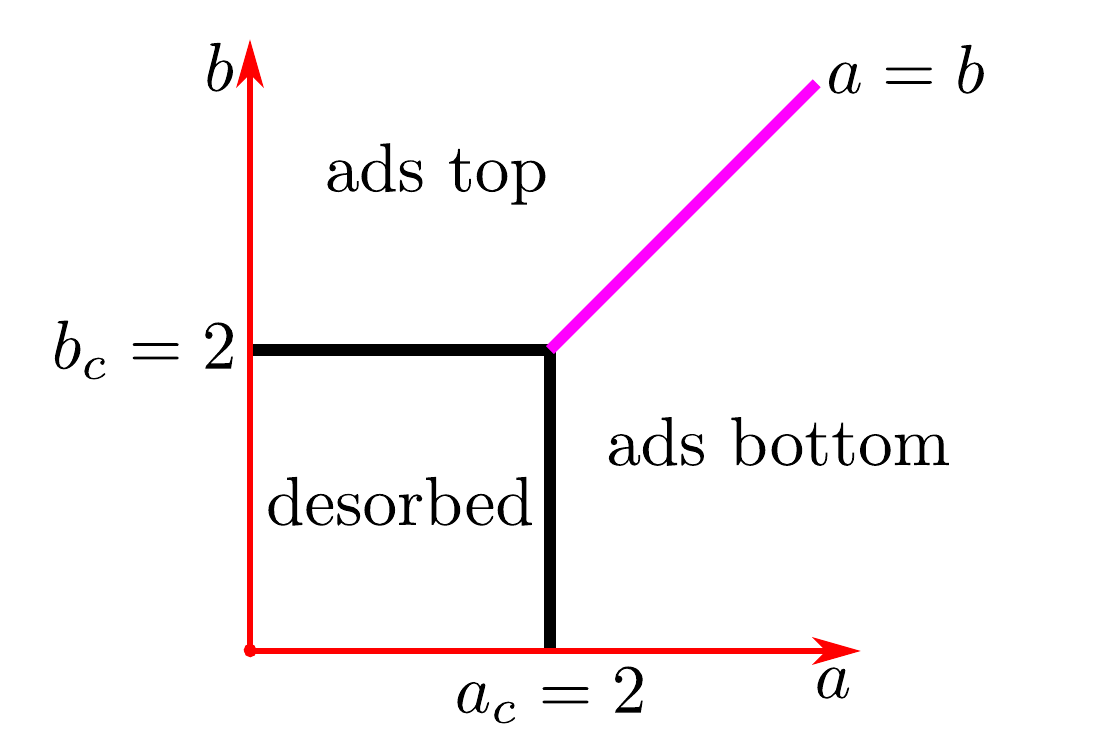}
\includegraphics[width=8cm]{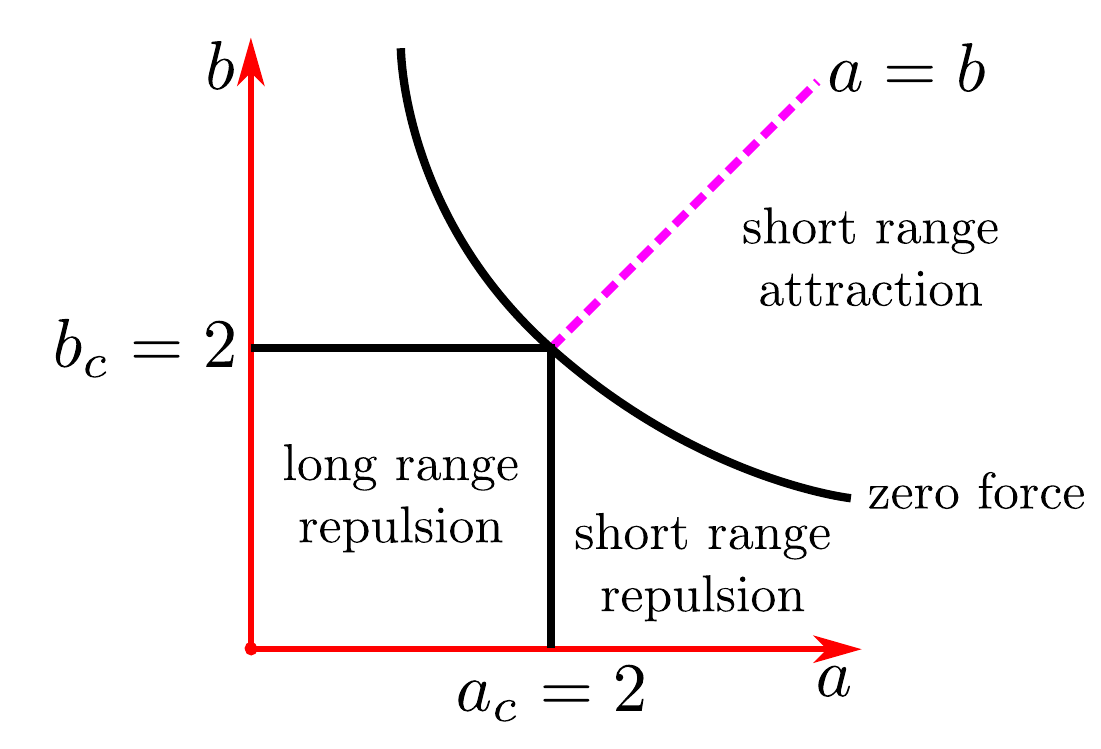}
\caption{(left) {Phase diagram of the infinite strip for a single walk. 
There are three phases: desorbed, adsorbed onto the bottom wall (ads 
bottom) and adsorbed onto the top (ads top).} (right) {A diagram of the 
regions of different types of effective force between the walls of a slit for a 
single Dyck path. Short range behaviour refers to exponential decay of the force 
with slit width while long range refers to a power law decay. The zero force 
curve is given by $ab=a+b$. On the dashed line there is a 
singular change of behaviour of the force.} }
\label{phase-force-diagram-single} 
\end{center}
\end{figure}

It was shown in \cite{brak2005a-:a} that 
\begin{align}
 \kappa^{single}_{inf-slit}(a,b) &=
\begin{cases}
\log \left(2\right) & 
\mbox{ if } a,b
        \leq 2 \\ 
\log\left(\frac{a}{\sqrt{a-1}}\right) & \mbox{ if } a > 2 \mbox{ and }
        a>b \\ 
\log\left(\frac{b}{\sqrt{b-1}}\right) & \mbox{ otherwise.} 
\end{cases}
\label{inf-stlit-free-energy-single}
\end{align}
For small $a$ and $b$ the walk is desorbed from both walls, while the large $a$ 
and $b$ phases are characterised by the order parameter of the thermodynamic 
density of visits to the bottom and top walls respectively. Correspondingly, 
there are 3 phase transition lines. The first two are given by $b=2$ for $0\leq 
a \leq 2$ and $a=2$ for $0\leq b \leq 2$. These lines separate the desorbed 
phase from the two adsorbed phases and are lines of second order transitions of 
the same nature as the one found in the half-plane model. There is also a first 
order transition for $a=b >2$ where the density of visits to each of the
walls jumps discontinuously  on crossing the boundary non-tangentially (see Figure~\ref{phase-force-diagram-single} (left)).

For finite widths the effective force between the walls, induced by the polymer,
was defined \cite{brak2005a-:a} as
\begin{align}
\mathcal{F}(a,b;w) &=   \kappa(a,b;w) - \kappa(a,b;w-1)\,.
\end{align}
For large $w$ it was found that the sign and length scale of the force depended on 
the values of $a$ and $b$ and that it was more refined that simply following the phase 
diagram (see Figure~\ref{phase-force-diagram-single}). 

The regions of the plane which gave different asymptotic expressions
for $\kappa$ and hence different phases for the infinite slit clearly
also give different force behaviours. For the square $0\leq a,b\leq 2$
the force is repulsive and decays as a power law (ie \ it is \emph{long-ranged}) 
while outside this square the force decays exponentially and so is
\emph{short-ranged}. This change coincides with the phase boundary of
the infinite slit phase diagram. However, the special curve $ab=a+b$
is a line of zero force across which the force, while short-ranged on
either side (except at $(a,b)=(2,2)$), changes sign. Hence this curve
separates regions where the force is attractive (to the right of the
curve) and repulsive to the left of the curve. The line $a=b$ for
$a>2$ is also special and, while the force is always
short-ranged and attractive, the range of the force on the line is
discontinuous and twice the size on this line than close by. All these
features leads to a \emph{force diagram} that encapsulates these
features (see Figure~\ref{phase-force-diagram-single} (right)). It should be recalled here that the behaviour 
of the directed system described above has been shown to be a faithful 
representation of the more general undirected self-avoiding walk model \cite{janse2005a-:a, martin2007a-:a}.

It is not unreasonable to speculate that the inequality of the infinite-slit and half plane limits, and more generally the resultant force diagram may be dependent on the particular single walk model chosen where the polymer was tethered to the bottom wall. There is no natural single walk model with fixed ends that can circumvent this restriction sensibly. One is therefore led to  consider  models of multiple walks in a slit where walks can be tethered to both walls. In fact a related generalisation has already been considered by
Alvarez \emph{et al.} \cite{alvarez2008self} where they studied  a model of 
self-avoiding polygons confined to a slit. The resulting force 
diagram is quite different from the single-walk diagram shown in 
Figure~\ref{phase-force-diagram-single} (left). 

In this paper we consider a directed walk model of two polymers confined between two 
walls with which the polymers interact, as in the single polymer model 
described above. In particular we fully analyse the infinite slit phase diagram and the large 
width force behaviour as a function of the interaction parameters. We show there are distinct differences from the single walk problem.

\section{Model}

We consider pairs of directed paths of equal length in a width $w$ strip
of the square lattice --- namely $\mathbb{Z}\times \{0,1,\dots, w\}$, taking
steps $(1,\pm1)$. These paths may touch (ie share edges and vertices) but not 
cross. We consider those pairs of paths whose initial vertices lie at at 
$(0,0)$ and $(0,w)$.
\begin{figure}[ht!]
\begin{center}
\includegraphics[width=10cm]{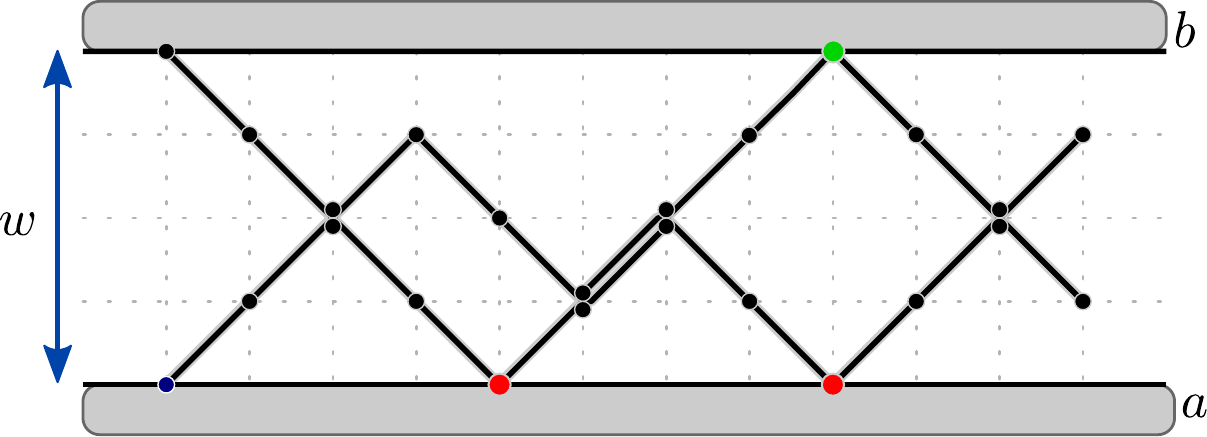}
\caption{ Two walks confined between two walls spaced $w$ lattice units 
apart. Each visit of the bottom  walk to the bottom wall contributes a Boltzmann 
weight $a$ and each visit of the top walk to the top wall contributes a 
Boltzmann weight $b$. For combinatorial reasons we do not weight the leftmost 
vertex of either walk.}
\label{twowalks} 
\end{center}
\end{figure}

Let $\varphi$ be such a pair of paths and define $|\varphi|$ to be the
length of the paths. If the width of the strip, $w$, is odd then the paths 
never share vertices and the combinatorics that follows is more complicated.
Because of this we only consider even widths. Note that this implies that the
distance between the endpoints of the paths is always even.

To complete our model let we add the energies
$-\varepsilon_a$ and $-\varepsilon_b$ for each visit of the walks to the bottom
and top walls respectively (aside from the leftmost vertex of each walk). The 
number of visits of the bottom walk to the bottom walk will be denoted 
$m_a(\varphi)$ while the number of visits of the top walk to the top wall will 
be denoted $m_b(\varphi)$ --- again excluding the leftmost vertex of each walk. 
The main model we discuss in the paper is based on pairs of walks, $\varphi$ ,
that finish with endpoints together at the same height. Define the corresponding
partition function to be
\begin{align}
Z_n(a,b;w) 
&= \sum_{\varphi} e^{(\varepsilon_a m_a(\varphi)+ \varepsilon_b 
m_b(\varphi) )/k_B T}
= \sum_{\varphi} a^{m_a(\varphi)} b^{m_b(\varphi)}\,,
\end{align}
where $T$ is the temperature, $k_B$ the Boltzmann constant 
and $a=e^{\varepsilon_a/k_B T}$ and $b=e^{\varepsilon_b/k_B T}$ are the 
Boltzmann weights associated with visits. The thermodynamic reduced free energy 
at finite width is given in the usual fashion as
\begin{align}
\kappa(a,b;w) &= \lim_{n \rightarrow \infty} \frac{1}{n}
\log\left(Z_n(w) \right).
\end{align}

Because the model at finite $w$ is essentially one-dimensional, the free
energy is an analytic function of $a$ and $b$ and no thermodynamic phase
transitions occur \cite{landau1980a-a}. As noted above, the \emph{infinite slit 
limit} for the single walk model does display singular behaviour and so we 
consider the same limit for this model. The \emph{infinite slit free energy} 
for the two walk model is found analogously by
\begin{align}
\kappa_{inf-slit}(a,b) &= \lim_{w\rightarrow\infty} \kappa(a,b;w)
=\lim_{w\rightarrow\infty} \lim_{n\rightarrow\infty}  
\frac{1}{n}  \log Z_n(a,b;w).
\end{align}
Motivated by the single walk problem, we see that the above quantity could be 
different when the order of limits is swapped. Since we have defined the model 
so that walks start on opposite walls, when the width is taken to infinity 
before the length, the system separates into two half-planes. Consequently we 
refer to this limit as the \emph{double half-plane limit} and so define 
\begin{equation}
 \kappa_{double-half-plane}(a,b) = \lim_{n\rightarrow\infty} 
\lim_{w\rightarrow\infty}  \frac{1}{n}  \log Z_n(a,b;w).
 \end{equation}
Since the system separates into two half-planes we have
\begin{equation}
     \kappa_{double-half-plane}(a,b) =
\kappa^{single}_{half-plane}(a) +\kappa^{single}_{half-plane}(b).
\label{double-half-plane-free-energy}
\end{equation}

Motivated by the single walk model, we consider the effective force applied to 
the walls by the polymers
\begin{align}
\mathcal{F}_n &=  \frac{1}{n} \left[ \log(Z_n(w)) -  \log(Z_n(w-2) ) 
\right],
\end{align}
with a thermodynamic limit of
\begin{align}
\mathcal{F}(a,b;w) &=   \kappa(a,b;w) - \kappa(a,b;w-2) .
\end{align}
Note that we will consider only systems of even width and hence we had to 
modify the single walk definition.

Given that the double half-plane limit is known from the discussion above, we 
shall concentrate on the infinite slit limit. In this limit, the free energy 
does not depend on where the walks end. It turns out that the combinatorics of 
the model in which the walks end together are easier. Accordingly we study
the generating function 
\begin{align}
G(a,b;z) &= \sum_{n=0}^\infty Z_n(w) z^n.
\end{align}
where the partition function now counts only those walks which end together. 
The radius of convergence of the generating function $z_c(a,b;w)$ is directly 
related to the free energy via
\begin{align}
\kappa(a,b;w) &= -\log\left(z_c(a,b;w)\right).
\end{align}

\section{Functional Equations}
\begin{figure}[h]
\begin{center}
 \includegraphics[height=3cm]{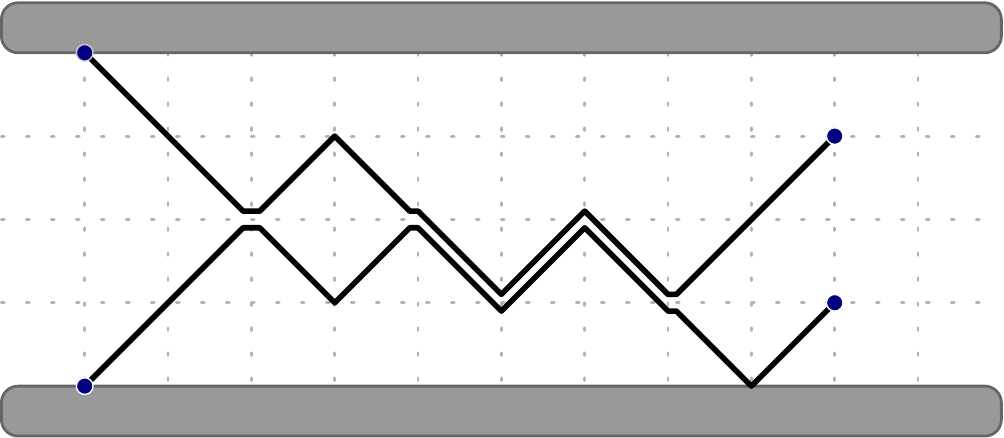}
\end{center}
\caption{We form the generating function of all pairs of paths that start 
in both surfaces and end anywhere according to their length, and distances of 
the endpoints from the surfaces. The path depicted contributes $z^9 r^1 s^1$ 
to the generating function.}
\label{fig feqn1}
\end{figure}

Though we are primarily interested in the behaviour of pairs of paths that
share their final vertices, we will need to define the generating function of 
more general pairs of paths with no restrictions on their endpoints. Define
$d_u( \varphi)$ to be the distance from the endpoint of the upper 
path to the top of the strip. Similarly define $d_\ell(\varphi)$ to 
be the
distance of the endpoint of the lower path to the bottom of the strip.
\subsection{Without interactions}
Let us first consider the case when $a=b=1$. We construct the generating 
function
\begin{align}
  F(r,s;z) \equiv F(r,s) &= \sum_{\varphi \in paths}
  z^{|\varphi|} r^{d_\ell( \varphi )} s^{d_u( \varphi )},
\end{align}
where $r,s$ are conjugate to the distances of the endpoints to either boundary
and $z$ is conjugate to length. See Figure~\ref{fig feqn1}. In order to
construct a functional equation satisfied by this generating function we also
need to define the generating function of those paths whose final vertices
touch.
\begin{align}
  r^w F_d(s/r;z) \equiv  r^w F_d(s/r)
  &= \sum_{h=0}^w s^h r^{w-h} \cdot \left[s^h r^{w-h}\right]\left\{ F(s,r) 
\right\}
\end{align}
where we have used $\left[s^i r^k\right]\left\{F(s,r)\right\}$ to denote the 
coefficient of $s^i r^k$ in the generating function $F(s,r)$. The generating 
function $G(1,1;z) = F_d(1;z)$. Also note that since the problem is vertically
symmetric, we have $F(r,s) \equiv F(s,r)$ and $r^w F_d(s/r) = s^w F_d(r/s)$. 
Further note that $\left[s^i r^k\right]\left\{F(s,r)\right\}$ is zero whenever 
$i-k$ is not even. 

One can construct all pairs of paths using a column-by-column construction 
whose details we give below. Translating the construction into its action on 
the generating functions gives the following functional equation
\begin{multline}
\label{eqn:nonint}
  F(r,s) = 1 + z\left(s+\frac{1}{s} \right) \left(r + \frac{1}{r} \right) \cdot 
F(r,s) \\
     - \frac{z}{r} \left(s + \frac{1}{s} \right) \cdot F(0,s)
     - \frac{z}{s} \left(r + \frac{1}{r} \right) \cdot F(r,0)
     + \frac{z}{sr} \cdot F(0,0) \\
     - z sr \cdot s^w F_d(r/s).
\end{multline}
We now explain each of the terms in this equation. The trivial pair of paths 
consists of two isolated vertices at $(0,0)$ and $(0,w)$. This gives the 
initial $1$ in the right-hand side of the above functional equation. Note that 
to enumerate directed polygons in the strip we may replace the above with all 
pairs of vertices lying at the same vertical ordinate; this
would replace $1$ with $\sum_{k=0}^w r^k s^{w-k} = (s^{w+1}-r^{w+1})/(s-r)$.
\begin{figure}[h!]
\begin{center}
 \includegraphics[height=3cm]{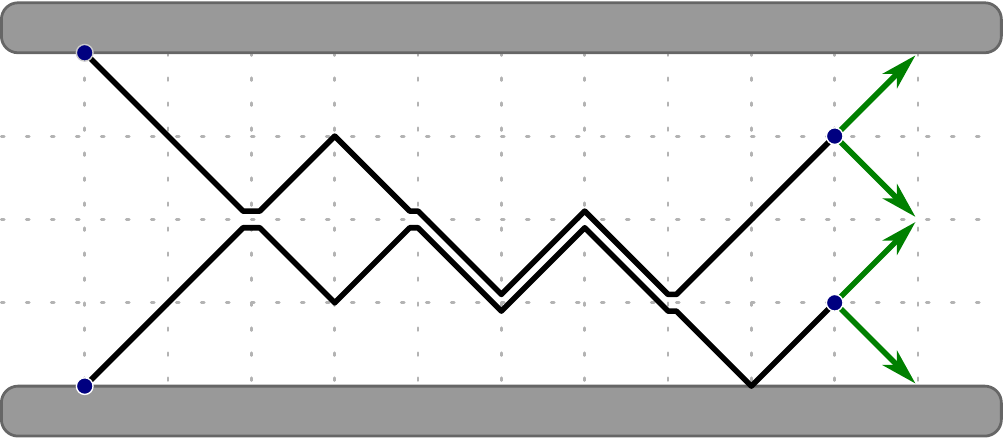}
\end{center}
  \caption{Every pair of paths can be continued by appending directed steps to
their endpoints as shown. While there are at most 4 possible combinations,
depending on the distance from boundaries, some combinations will be forbidden.}
  \label{fig feqn2}
\end{figure}

See Figure~\ref{fig feqn2}. When the endpoints are away from the boundaries, 
every pair of paths may be continued by appending directed steps in four 
different ways. Since each of these steps either increases or decreases the 
distance of the endpoint from the boundary the result is
\begin{align}
  z\left(s+\frac{1}{s} \right) \left(r + \frac{1}{r} \right) \cdot F(r,s).
\end{align}

\begin{figure}[h!]
 \begin{center}
 \includegraphics[height=3cm]{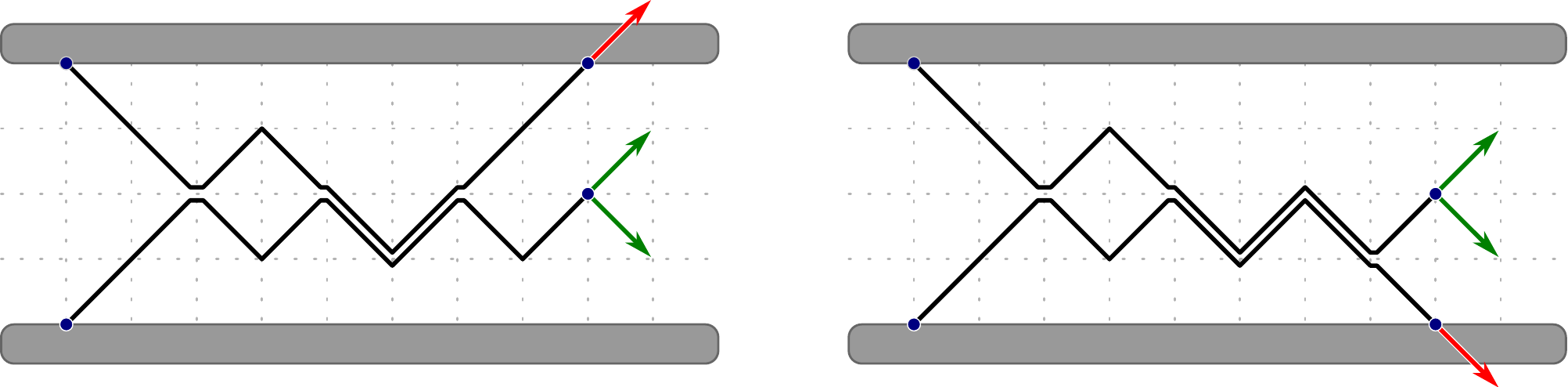}
\end{center}
  \caption{When the endpoints of the walks are close to the boundaries one must
take care to subtract off the contributions of the configurations that
step outside the strip as depicted here.}
  \label{fig feqn3}
\end{figure}

See Figure~\ref{fig feqn3}. When the endpoints are close to the boundaries or
each other, then appending steps as described above may result in paths that
either step outside the strip or cross each other. If the endpoint of the upper
path lies on the boundary then one cannot append a $(1,1)$ step to that path.
Such configurations are counted by 
\begin{align}
  \frac{z}{s}\left(r+\frac{1}{r} \right) \cdot \left[s^0\right] F(r,s)
  & \equiv \frac{z}{s}\left(r+\frac{1}{r} \right) F(r,0).
\end{align}
Similarly if the endpoint of the lower path lies on the boundary then one cannot 
append a $(1,-1)$ step to that path:
\begin{align}
  \frac{z}{r}\left(s+\frac{1}{s} \right) \cdot \left[r^0\right] F(r,s)
  & \equiv \frac{z}{r}\left(s+\frac{1}{s} \right) F(0,s).
\end{align}
\begin{figure}[h!]
\begin{center}
 \includegraphics[height=3cm]{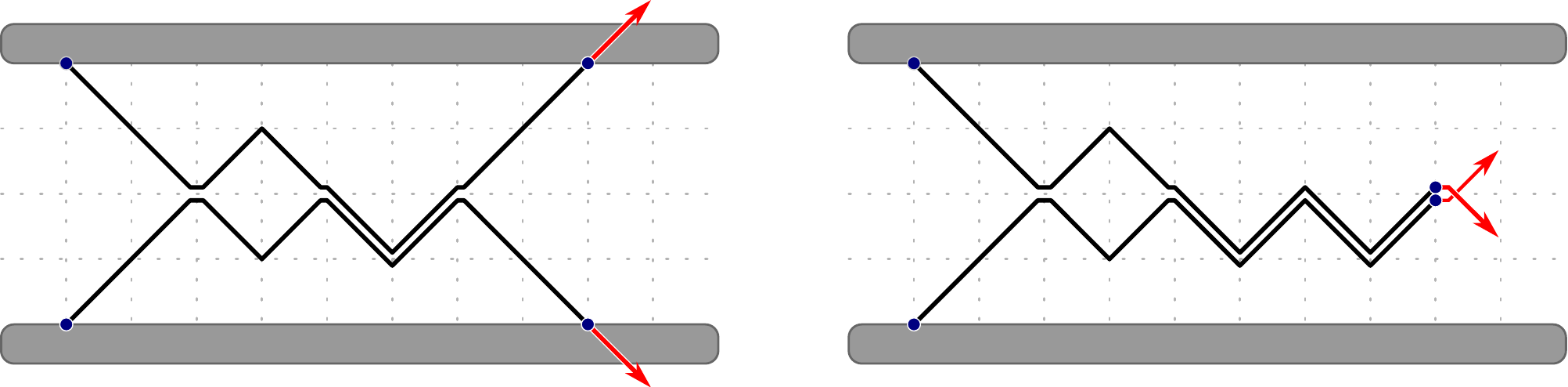}
\end{center}
  \caption{(left) When removing the contributions of paths that step
outside the strip, we over-correct by twice removing those configurations
in which both paths step outside the strip simultaneously. (right) When the
endpoints of the paths are close together we must remove the contribution of
paths that cross each other.}
  \label{fig feqn4}
\end{figure}

We correct the enumeration by subtracting both of these contributions. In so
doing we over-correct by subtracting twice the contribution of paths whose
endpoints lie on opposite boundaries (see Figure~\ref{fig feqn4}(left)). Thus
we add back in
\begin{align}
  \frac{z}{sr}\cdot \left[s^0r^0\right] F(r,s)
  & \equiv \frac{z}{sr} F(0,0)
\end{align}
Finally, we must also remove the contribution of those paths whose
endpoints cross. This happens when we take a path whose endpoints lie together
and attempt to append an upward step to the lower path and a downward step to
the upper path (see Figure~\ref{fig feqn4}(right)). So we must subtract
\begin{align}
  \label{eqn fd symm}
     z sr \cdot r^w F_d(s/r) & \equiv zsr \cdot s^w F_d(r/s),
\end{align}
where this equivalence comes from the vertical symmetry of the model without
interactions.

\subsection{Interacting model}
\label{sec simple int}
We now add boundary interactions to this model. We weight each pair of paths 
according to the number of contacts the upper (lower) path has with the upper 
(lower) boundary excluding their leftmost vertices. Recall that $a$ is conjugate to the 
number of contacts between the lower path and the boundary and similarly 
$b$ is conjugate to the number of contacts between the upper path and the 
boundary. Thus our generating functions $F$ and $F_d$ become functions of $a,b$ 
in addition to $r,s,z$; as above we will typically write these as
\begin{align}
  F(r,s;a,b;z) \equiv F(r,s) &&\text{and } &&  F_d(s/r;a,b;z) \equiv F_d(s/r).
\end{align}

\begin{figure}[h!]
\begin{center}
 \includegraphics[height=3.6cm]{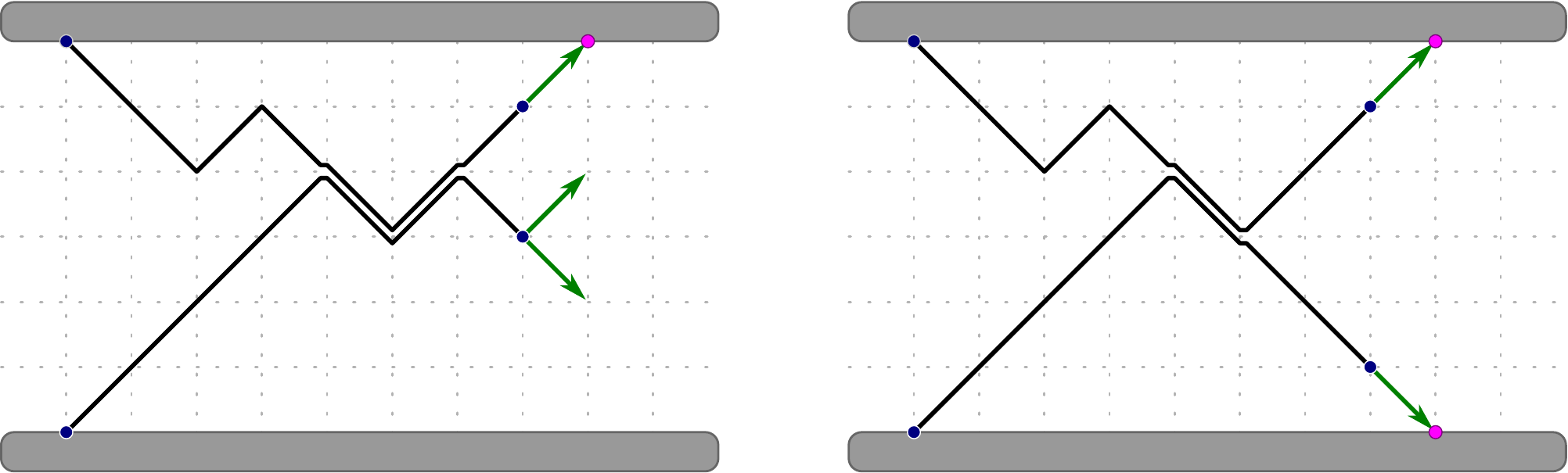}
\end{center}
\caption{Interactions with the boundary are produced when one or both paths
steps from distance one onto the boundary.}
\label{fig feqn5}
\end{figure}

We now modify the above construction by noting that a contact between the upper 
path and its boundary is created when an upward step is appended to a path lying 
1 step from the boundary (see Figure~\ref{fig feqn5}). Thus we add
\begin{align}
  zb\left(r + \frac{1}{r} \right) \left[s^1 \right] \left\{F(r,s) \right\}.
\end{align}
However these configurations have already been enumerated with incorrect weight, 
so we must also subtract
\begin{align}
   z \left(r + \frac{1}{r} \right) \left[s^1 \right] \left\{F(r,s) \right\}.
\end{align}
Thus we arrive at
\begin{align}
  z(b-1) \left(r + \frac{1}{r} \right) \left[s^1 \right] \left\{F(r,s) \right\}.
\end{align}
And similarly, by considering contacts between the lower path and the lower 
boundary we obtain
\begin{align}
  z(a-1) \left(s + \frac{1}{s} \right) \left[r^1 \right] \left\{F(r,s) \right\}.
\end{align}
Again we find that these terms over-correct and we must consider those 
configurations in which contacts with the upper and lower boundaries are created 
at the same time.
\begin{align}
  z(a-1)(b-1) \left[s^1 r^1 \right] \left\{F(r,s) \right\}.
\end{align}
So finally we have the following functional equation for $F(r,s;a,b;z) \equiv 
F(r,s)$.
\begin{align}
  F(r,s)
  =& 1 + z\left(s+\frac{1}{s} \right) \left(r + \frac{1}{r} \right) \cdot F(r,s) 
\nn
  &- \frac{z}{r} \left(s + \frac{1}{s} \right) \cdot F(0,s)
     - \frac{z}{s} \left(r + \frac{1}{r} \right) \cdot F(r,0)
     + \frac{z}{sr} \cdot F(0,0)- z sr \cdot s^w F_d(r/s) \nn
 &+z(b-1)\left(r + \frac{1}{r} \right)\left[s^1\right]\left\{F(r,s)\right\}
  +z(a-1)\left(s + \frac{1}{s} \right)\left[r^1\right]\left\{F(r,s)\right\}\nn
 &+z(a-1)(b-1)\left[s^1 r^1 \right] \left\{F(r,s) \right\}.
\end{align}

We can now further simplify this equation by rewriting
$\left[s^1\right]\left\{F(r,s)\right\}, \left[r^1\right]\left\{F(r,s)\right\}$ 
and
$\left[s^1 r^1 \right] \left\{F(r,s) \right\}$ in terms of $F(r,0), F(0,s)$ and 
$F(0,0)$.

Extracting the coefficient of $s^0r^0$ in the above equation gives
\begin{align}
  F(0,0) =
  &1+z\left( 1 + (b-1)+(a-1) +(a-1)(b-1) \right)\left[s^1 
r^1\right]\left\{F(r,s)\right\}
  \nn
  =& 1 + zab \left[s^1r^1 \right] \left\{ F(s,r) \right\}.
\end{align}
This has a simple combinatorial interpretation; any path with endpoints 
ending in each surface must either be trivial or can be constructed from a 
shorter path whose endpoints end a single unit from each boundary.

Similarly, extracting the coefficient of $s^0$ in the above gives
\begin{align}
  F(r,0) &= 1
  + z \left(r+\frac{1}{r}\right)\left[s^1\right]\left\{F(r,s)\right\}
  -\frac{z}{r} \left[s^1r^0\right]\left\{ F(s,r) \right\} \nn
  &+z(b-1)\left(r + \frac{1}{r} \right)\left[s^1\right]\left\{F(r,s)\right\}
  +z(a-1)\left[r^1s^1\right]\left\{F(r,s)\right\}\nn
  &+ z(a-1)(b-1) \left[s^1 r^1 \right]\left\{F(r,s)\right\} \nn
  &= 1 + zb \left(r+\frac{1}{r}\right)\left[s^1\right]\left\{F(r,s)\right\}
  + zb(a-1) \left[s^1 r^1 \right]\left\{F(r,s)\right\} 
\end{align}
and similarly
\begin{align}
  F(0,s)
  &= 1 + za \left(r+\frac{1}{r}\right)\left[r^1\right]\left\{F(r,s)\right\}
  + za(b-1) \left[s^1 r^1 \right]\left\{F(r,s)\right\}. 
\end{align}
This gives three linear equations and we may solve them to obtain
\begin{align}
z\left[s^1 r^1 \right] \left\{F(r,s) \right\}
&= \frac{F(0,0)-1}{ab} \\
z\left(s + \frac{1}{s} \right) \left[r^1\right]\left\{F(r,s)\right\}
&= -\frac{1 + F(r,0) + (b-1)F(0,0)}{ab}\\
z \left(r + \frac{1}{r} \right) \left[s^1\right]\left\{F(r,s)\right\}
&= -\frac{1 - F(0,s) + (a-1)F(0,0)}{ab}.
\end{align}
Substituting these into the original interactions equation gives us
\begin{align}
\label{eqn:functional eqn}
  F(r,s) =&
  \frac{1}{ab}
  + z\left(s+\frac{1}{s} \right) \left(r + \frac{1}{r} \right) \cdot F(r,s) 
  - z sr \cdot s^w F_d(r/s)\nn
  &+ A(r,s) F(0,s) + B(r,s) F(r,0) + C(r,s) F(0,0),
  \intertext{where}
  A(r,s) &= 1-\frac{1}{b} - \frac{z(r+1/r)}{s},\nn
  B(r,s) &= 1-\frac{1}{a} - \frac{z(s+1/s)}{r},\nn
  C(r,s) &= \frac{z}{sr}-\left(1-\frac{1}{a}\right)\left(1-\frac{1}{b}\right).
\end{align}
From this equation we can recover $G(a,b;z)=F_d(1;a,b;z)$. In the 
following section we do not solve explicitly for $F_d$, however we are able to 
determine its singularities and so its asymptotic behaviour. 

\section{Solution of Functional Equations}

At this point, we define $v = w/2$ as is the more natural parameter in what 
follows. Rather than solving the full model directly, we first examine the 
special cases of $a=b=1$ and $a=b$.
\subsection{Without Interactions }

We start by collecting the $F(r,s)$ terms in equation~\Ref{eqn:nonint} to get
\begin{multline}
  \left(1- z\left(s+\frac{1}{s} \right) \left(r + \frac{1}{r}
\right)\right)F(r,s) = 1 - \frac{z}{r} \left(s + \frac{1}{s} \right) \cdot
F(0,s) \\
  - \frac{z}{s} \left(r + \frac{1}{r} \right) \cdot F(r,0)  + \frac{z}{sr} \cdot
F(0,0) - z sr \cdot s^{2v} F_d(r/s).
\end{multline}

The coefficient of $F(r,s)$ is called the kernel $K(r,s;z) \equiv K(r,s)$ and
its symmetries play a key role in the solution
\begin{align}
K(r,s) &= 1- z\left(s+\frac{1}{s} \right) \left(r + \frac{1}{r} \right).
\end{align}

We use the kernel method which exploits the symmetries of the kernel to
remove boundary terms in the functional equation (see \cite{bousquet2010walks} 
for a thorough description of the kernel method).
The kernel is symmetric under the following operations
\begin{align}
(r,s) &\mapsto \left(\frac{1}{r},s\right) &
(r,s) &\mapsto \left(r,\frac{1}{s}\right) &
(r,s) &\mapsto \left(s,r\right).
\end{align}
To more be precise, we use the above symmetries to construct the following 
equations
\begin{subequations}
\footnotesize
\begin{align}
\label{eqn symm1}
K(r,s) \cdot F(r,s) 
&= 1-\frac{z}{s}\left(r+\frac{1}{r} \right)F(r,0) -
\frac{z}{r}\left(s+\frac{1}{s}\right)F(0,s) + \frac{z}{sr}F(0,0) -
zrs^{{2v}+1}F_d\left(\frac{r}{s}\right)\\
\label{eqn symm2}
K\left(\frac{1}{r},s \right) \cdot F\left(\frac{1}{r},s \right)
&= 1-\frac{z}{s}\left(r+\frac{1}{r} \right)F\left(\frac{1}{r},0\right) -
zr\left(s+\frac{1}{s}\right)F(0,s) + \frac{zr}{s}F(0,0) - \frac{zs^{{2v}+1}}{r}
F_d\left(\frac{1}{rs}\right)\\
\label{eqn symm3}
K\left(r,\frac{1}{s} \right) \cdot F\left(r,\frac{1}{s} \right)
&= 1-zs\left(r+\frac{1}{r} \right)F\left(r,0\right) -
\frac{z}{r}\left(s+\frac{1}{s}\right)F\left(0	,\frac{1}{s} \right) +
\frac{zs}{r}F(0,0) - \frac{zr}{s^{{2v}+1}} F_d\left(rs\right)\\
\label{eqn symm4}
K\left(\frac{1}{r},\frac{1}{s} \right) \cdot F\left(\frac{1}{r},\frac{1}{s}
\right)
&= 1-zs\left(r+\frac{1}{r} \right)F\left(\frac{1}{r},0\right) -
zr\left(s+\frac{1}{s}\right)F\left(0	,\frac{1}{s} \right) + zrsF(0,0) -
\frac{z}{rs^{{2v}+1}} F_d\left(\frac{s}{r}\right).
\end{align}
\end{subequations}

We can eliminate the boundary terms by taking the appropriate alternating sum of
the above equations:
\begin{align}
rs\cdot \text{Eqn}(\ref{eqn symm1}) 
- \frac{s}{r}\cdot \text{Eqn}(\ref{eqn symm2}) 
- \frac{r}{s} \cdot \text{Eqn}(\ref{eqn symm3}) 
+ \frac{1}{rs}\cdot \text{Eqn}(\ref{eqn symm4}).
\end{align}
This is similar to the ``orbit-sum'' discussed in \cite{bousquet2010walks, bousquet2010expected}.

Since the kernel is the same in all of the above equations we obtain
\begin{multline}
 K(r,s) \cdot \left(\text{Sum of $F$} \right) = 
      \frac{(s-1)(s+1)(r-1)(r+1)}{rs}\\
     +\frac{zs^{{2v}+2}}{r^2}F_d\left(\frac{1}{rs}\right)
+\frac{zr^2}{s^{{2v}+2}}F_d\left(rs\right)\\
     - z{s^{{2v}+2}}{r^2}F_d\left(\frac{r}{s}\right) -
\frac{z}{r^2s^{{2v}+2}}F_d\left(\frac{s}{r}\right).
\end{multline}

The symmetry of $F_d$ described by equation~\Ref{eqn fd symm} comes from the
vertical symmetry of the model; it can be extended to give
\begin{align}
zr^{{2v}+1}s F_d\left(\frac{s}{r}\right) &\equiv zrs^{{2v}+1}
F_d\left(\frac{r}{s}\right) & 
zr^{{2v}+1}s^{{2v}+1} F_d\left(\frac{1}{rs}\right) &\equiv 
F_d\left(\frac{r}{s}\right).
\end{align}

These relations will then simplify the functional equation further:
\begin{multline}
 K(r,s) \cdot \left(\text{Sum of $F$}\right) = 
      \frac{(s-1)(s+1)(r-1)(r+1)}{rs}\\
     +\frac{z\left(r^{{2v}+4} +
s^{{2v}+4}\right)}{s^{{2v}+2}r^{{2v}+2}}F_d\left(rs\right) -
\frac{z\left({r^{{2v}+4}s^{{2v}+4}+1}\right)}{r^{{2v}+2}s^2}F_d\left(\frac{r}{s}
\right).
\end{multline}

We can now remove the left hand side of the equation by choosing values of $r$
and $s$ that set the kernel to zero. That is, $K(\hat{r}, \hat{s}) = 0$; this
also gives $z^{-1} = \left(\hat{s}+\frac{1}{\hat{s}} \right) \left(\hat{r} + 
\frac{1}{\hat{r}} \right)$. Making this substitution gives
\begin{multline}
0 = \frac{(\hat{s}-1)(\hat{s}+1)(\hat{r}-1)(\hat{r}+1)}{\hat{r}\hat{s}} 
+\frac{\left(\hat{r}^{{2v}+4} +
\hat{s}^{{2v}+4}\right)}{\hat{s}^{{2v}+1}\hat{r}^{{2v}+1}\left(\hat{r}
^2+1\right)\left(\hat{s}^2+1\right)}F_d\left(\hat{r}\hat{s}\right)\\
 -
\frac{\left({\hat{r}^{{2v}+4}\hat{s}^{{2v}+4}+1}\right)}{\hat{r}^{{2v}+1}\hat{s}
\left(\hat{r}^2+1\right)\left(\hat{s}^2+1\right)
}F_d\left(\frac{\hat{r}}{\hat{s}}\right).
\end{multline}
By eliminating denominators, we obtain
\begin{multline}
\label{eqn:nointfunct}
0 =
(\hat{s}-1)(\hat{s}+1)(\hat{r}-1)(\hat{r}+1)\left(\hat{r}^2+1\right)\left(\hat{s
}^2+1\right) \hat{r}^{2v} \hat{s}^{2v}\\
     +\left(\hat{r}^{{2v}+4} + \hat{s}^{{2v}+4}\right)
F_d\left(\hat{r}\hat{s}\right) - \hat{s}^{2v}
\left({\hat{r}^{{2v}+4}\hat{s}^{{2v}+4}+1}\right)F_d\left(\frac{\hat{r}}{\hat{s}
}\right).
\end{multline}

We now apply a similar argument used by Bousquet-M{\'e}lou in 
\cite{bousquet2010expected} to determine the singularities of $F_d$. Set 
$\hat{r} = q\hat{s}$ for a root of unity $q \neq -1$ such that $q^{{2v}+4} = 
-1$. More precisely, we choose $\hat{s}$ as a solution to $K(qs,s) = 0$. The 
above equation then reduces to
\begin{multline}
0 = (\hat{s}^4q^4 -1)(\hat{s}^4-1) (\hat{s}q)^{2v} \hat{s}^{2v}\\
     +\hat{s}^{{2v}+4}\left(q^{{2v}+4} + 1\right) F_d\left(q\hat{s}^2\right) -
\hat{s}^{2v} \left({q^{{2v}+4}\hat{s}^{{4v}+8}+1}\right)F_d\left(q\right).
\end{multline}
Since $q^{{2v}+4} = -1$, the second term drops out and we can find an explicit
equation for $F_d(x)$ at the roots of unity $q$.
\begin{align}
F_d(q) &= \frac{(\hat{s}^4q^4 -1)(\hat{s}^4-1) (\hat{s}q)^{2v} }{1-
\hat{s}^{4v+8}}.
\end{align}
Since the kernel $K(q\hat{s},\hat{s})$ is quadratic in $\hat{s}^2$, this implies
symmetric functions in $\hat{s}^2$ will also be rational in $z$. By rewriting
$F_d(q)$ as
\begin{align}
\label{eqn:fdqk}
F_d(q) &= \frac{(\hat{s}^2q^2
-\frac{1}{\hat{s}^2q^2})(\hat{s}^2-\frac{1}{\hat{s}^2}) q^{{2v}+2}
}{\hat{s}^{-({2v}+4)}- \hat{s}^{{2v}+4}},
\end{align}
we can see that it must also be rational in $z$. Of course, one can see much 
more directly that $F_d$ must be a rational function of $z$ since it can be 
translated into a problem of counting paths via a finite transfer matrix (see, 
for example, Chapter~V of \cite{flajolet2009analytic}).

The construction of $F_d(x)$ ensures that it is a polynomial in $x$ of degree
${2v}$. Thus, we can obtain the full $F_d(x)$ by using Lagrange polynomial
interpolation and the known points of $F_d(q)$ (we follow the method in 
\cite{bousquet2010expected}). By taking a set of $\{q_k\}$ such that $q_k 
^{{2v}+4} = -1$ with $q_i \neq -1$ for any $i$ and
making the substitutions, we get

\begin{align}
F_d(x) = \sum_{j=0}^{{2v}} F_d(q_j) \prod_{\substack{0 \leq m \leq {2v} \\ m
\neq j}} \frac{x - q_m}{q_j - q_m}.
\end{align}

Note that no term in the product contributes any singularities in $z$. Thus
$F_d(x)$ being singular implies at least one $F_d(q_k)$ is also singular. By
equation~\Ref{eqn:fdqk}, we can see that $F_d(q)$ will be singular when 
$\hat{s}$ (and hence $\hat{r}$) is a $(4v+8)$-th root of unity. Combining with 
the kernel, a superset of singularities can obtained by various choices of 
$k,j$:
\begin{align}
\label{eqn:a1b1soln}
z_{j,k} = \frac{1}{\left(\hat{r} + \frac{1}{\hat{r}}\right)
\left(\hat{s} + \frac{1}{\hat{s}}\right)} =
\frac{1}{4 \cos\left(\frac{\pi j}{2v+4}\right)\cos\left(\frac{\pi
k}{2v+4}\right)}.
\end{align}
Note that since $\hat{r} = q\hat{s}$ with $q^{{2v}+4} = -1$, we do not have 
$j=k$ in the above and so the dominant singularity is obtained when $j=1,k=2$ 
(or vice-versa).
\subsection{With Equal Interactions $a=b$}

For this section, we follow the same argument however the details become more 
complicated due to the boundary terms. We start by arranging 
equation~\Ref{eqn:functional eqn} to collect all $F(r,s)$ terms to obtain the 
equation
\begin{align}
  K(r,s)F(r,s) =&
  \frac{1}{ab} - z sr \cdot s^{2v} F_d(r/s)\nn
  &+ A(r,s) F(0,s) + B(r,s) F(r,0) + C(r,s) F(0,0),
  \intertext{where}
  A(r,s) &= 1-\frac{1}{a} - \frac{z(r+1/r)}{s},\nn
  B(r,s) &= 1-\frac{1}{a} - \frac{z(s+1/s)}{r},\nn
  C(r,s) &= \frac{z}{sr}-\left(1-\frac{1}{a}\right)^2
\end{align}
with the kernel 
\begin{equation}
K(r,s) = 1 - z\left(r + \frac{1}{r}\right)\left(s + \frac{1}{s}\right).
\end{equation}
Since the kernel is the same as that of the non-interacting case we can use the 
same symmetries and combine the four equations to eliminate the boundary terms 
$F(r,0), F\left(\frac{1}{r},0\right), F(0,s)$ and 
$F\left(0,\frac{1}{s}\right)$. This results in the following functional 
equation
\begin{multline}
K(r,s) \cdot\left(\text{linear combination of } F \right) =\\
rs^{{2v}+1}(s^2-1)(r^2-1)(a-1)^2 (r^2s^2z+r^2z-sr+s^2z+z)z \cdot F(0,0)\\
 +(sza+r^2sza+r-ra)(rsa-rs-s^2za-za)zs^{4v+3} \cdot 
F_d\left(\frac{1}{rs}\right)\\
 -(rsa-rs-s^2za-za)(za+r^2za-rsa+rs)z\cdot  F_d\left(\frac{s}{r}\right) \\
 +(sza+r^2sza+r-ra)(rs^2za+rza+s-sa)zr^3s^{4v+3}\cdot 
F_d\left(\frac{r}{s}\right)\\
 -(za+r^2za-rsa+rs)(rs^2za+rza+s-sa)zr^3 \cdot  F_d\left(rs\right)\\
-  rs^{{2v}+1}z^2(s^4-1)(r^4-1).
\end{multline}

Since the wall interaction is symmetric, we can again make use of the vertical
symmetry to eliminate $F_d\left(\frac{1}{rs}\right)$ and
$F_d\left(\frac{s}{r}\right)$ and give
\begin{multline}
K(r,s) \cdot\left(\text{linear combination of } F \right) \\
= L(r,s;a)\cdot
F(0,0) +M(r,s;a)\cdot F_d\left(\frac{r}{s}\right) \\
+N(r,s;a)\cdot F_d\left({r}{s}\right)-  rs^{{2v}+1}z^2(s^4-1)(r^4-1),
\end{multline}
where $L(r,s;a),M(r,s;a)$ and $N(r,s;a)$ are easily computed though complicated 
functions. As before, we pick values of $r$ and $s$ that set the kernel to
$0$. And since $K(\hat{r},\hat{s}) =0$ we can write $z = 
(\hat{s}+1/\hat{s})^{-1}(\hat{r}+1/\hat{r})^{-1}$ and so eliminate it from the
coefficients of the above equations. After clearing the denominators, we obtain
a functional equation with coefficients $\alpha, \beta$ and $\delta$ (again 
being easily computed, though complicated, functions)
\begin{align}
0 &= \alpha(r,s;a) \cdot F_d\left(\frac{\hat{r}}{\hat{s}}\right) 
+ \beta(r,s;a) \cdot F_d\left(\hat{r}s\right)+ \delta(r,s;a).
\end{align}
The coefficient $\delta$ is important in what follows, and so we state it 
explicitly
\begin{align}
\delta(r,s;a) &= r^{2v}s^{2v}(1-r^4)(1-s^4).
\end{align}
Note that if $r$ or $s$ are fourth roots of unity then $\delta=0$.

Unlike the $a=b=1$ case, there is no simple relation between $\hat{r}$
and $\hat{s}$ that will give us an explicit form for $F_d(x)$. However, we can
still extract the location of the singularities by solving when the
coefficients $\alpha$ and $\beta$ are simultaneously $0$ with $\delta \neq 0$.
Solving $\alpha=\beta=0$, we get
\begin{align}
\label{eqn:rssoln}
\hat{r}^{{2v}} &= \frac{ \hat{r}^2 (a-1)-1 }{\hat{r}^2(a-1-\hat{r}^2)} &
\hat{s}^{2v} &= - \frac{\hat{s}^2(a-1)-1 }{\hat{s}^2(a-1-\hat{s}^2)}
\intertext{or}
\hat{r}^{{2v}} &= - \frac{ \hat{r}^2 (a-1)-1 }{\hat{r}^2(a-1-\hat{r}^2)} &
\hat{s}^{2v} &= \frac{\hat{s}^2(a-1)-1 }{\hat{s}^2(a-1-\hat{s}^2)}.
\end{align}
Since the form of $\hat{r}$ and $\hat{s}$ is similar, we will concentrate on
finding the solutions of
\begin{align}
\hat{r}^{2v} = \frac{ \hat{r}^2 (a-1)-1 }{\hat{r}^2(a-1-\hat{r}^2)}
\end{align}
and from there, deducing solutions for $\hat{s}$.

By rearranging the equation, we get
\begin{equation}
\label{eqn:areduction}
a-1 = \frac{\hat{r}^{v+2} - \frac{1}{\hat{r}^{v+2}}} {\hat{r}^v -
\frac{1}{\hat{r}^v}}.
\end{equation}
Since $a$ is a positive real parameter, the right hand side must also be real.
The following theorem tells us that all solutions to this equation must lie
either on the unit circle or the real line.

\begin{thm}
\label{thm:rrootsloc}
The expression
\begin{equation}
\frac{\hat{r}^{v+2} - \frac{1}{\hat{r}^{v+2}}} {\hat{r}^v -
\frac{1}{\hat{r}^v}}
\end{equation}
is real if and only if $\hat{r} \in \mathbb{R}$ or if $|\hat{r}| = 1$. The
equivalent statement holds for $\hat{s}$.
\end{thm}
The proof of this statement is given in the appendix and is relatively 
straightforward though cumbersome.

We can further refine the above statement when $a \leq 2$ and in that case all 
the solutions lie on the unit circle. To do this, we use of Theorem~$1$ from
Lal{\'\i}n and Smyth \cite{lalin2013unimodularity}.

\begin{thm}[from \cite{lalin2013unimodularity}]
\label{thm:smalla}
Let $h(z)$ be a non-zero complex polynomial of degree $n$ having all its zeros
in the closed unit disc $|z| \leq 1$. Then for $d > n$ and any $\lambda$ on the
unit circle, the self inverse polynomial
\begin{equation}
P^{(\lambda)}(z) = z^{d-n} h(z) + \lambda z^n \bar{h}\left(\frac{1}{z}\right)
\end{equation}
has all its zeros on the unit circle.
\end{thm}

By rearranging equation~\Ref{eqn:rssoln}, we get
\begin{align}
0 &= \hat{r}^{2v+2}(a-1-\hat{r}^2) - (\hat{r}^2 (a-1)-1) \\
0 &= \hat{s}^{2v+2}(a-1-\hat{s}^2) + (\hat{s}^2(a-1)-1)
\end{align}
which is in the form given in the theorem with $n=2$, $h(z) = (a-1 - z^2)$ and
$\lambda = \pm1$. The zeros of $h(z)$ are given by
\begin{align}
z &= \pm \sqrt{a-1}.
\end{align}
Hence, the zeros of $h(z)$ will be inside the closed disc exactly when $a \leq
2$ and so we can apply the theorem.

We note that when $\hat{r}$ and $\hat{s}$ lie on the unit circle, the 
singularities of the generating function are of a similar form to that given 
in equation~\Ref{eqn:a1b1soln}. However, the angles are not simple functions of 
$w$. In Section~\ref{subs:case4} we give asymptotic expressions for the 
singularities.

\subsection{With Interactions, $a$, $b$ free}
We proceed via the same argument as per the previous sections. We start by
arranging equation~\Ref{eqn:functional eqn} to collect all $F(r,s)$ terms to
obtain the equation
\begin{align}
  K(r,s)F(r,s) =&
  \frac{1}{ab} - z sr \cdot s^{2v} F_d(r/s)\nn
  &+ A(r,s) F(0,s) + B(r,s) F(r,0) + C(r,s) F(0,0),
  \intertext{where}
  A(r,s) &= 1-\frac{1}{b} - \frac{z(r+1/r)}{s},\nn
  B(r,s) &= 1-\frac{1}{a} - \frac{z(s+1/s)}{r},\nn
  C(r,s) &= \frac{z}{sr}-\left(1-\frac{1}{a}\right)\left(1-\frac{1}{b}\right)
\end{align}
with the same kernel as before.

Again, use the symmetries of the kernel to construct 4 linear equations and 
then take linear combinations to eliminate the boundary terms $F(r,0),
F\left(\frac{1}{r},0\right)$, $F(0,s)$ and $F\left(0,\frac{1}{s}\right)$. This
results in the following functional equation
\begin{multline}
K(r,s) \cdot\left(\text{linear combination of } F \right)\\ =
rs^{{2v}+1}(s^2-1)(r^2-1)(a-1)(b-1) (r^2s^2z+r^2z-sr+s^2z+z)z \cdot F(0,0)\\
 +(szb+r^2szb+r-rb)(rsa-rs-s^2za-za)zs^{4v+3} \cdot 
F_d\left(\frac{1}{rs}\right)\\
 -(rsa-rs-s^2za-za)(zb+r^2zb-rsb+rs)z\cdot  F_d\left(\frac{s}{r}\right) \\
 +(szb+r^2szb+r-rb)(rs^2za+rza+s-sa)zr^3s^{4v+3}\cdot 
F_d\left(\frac{r}{s}\right)\\
 -(zb+r^2zb-rsb+rs)(rs^2za+rza+s-sa)zr^3 \cdot  F_d\left(rs\right)\\
 -rs^{{2v}+1}z^2(s^4-1)(r^4-1).
\end{multline}

Unlike the previous case, the wall interactions are no longer symmetric and
hence we cannot apply the vertical symmetry. However, we can pick values
$\hat{r}$ and $\hat{s}$ that sets the kernel $K(\hat{r},\hat{s}) = 0$ and
eliminate $z$ from the equation. Making this substitution, we get
\begin{multline}
\label{eqn:abfuncteqn}
0 = \hat{s}^{4v+2} (b-1-\hat{s}^2)(\hat{r}^2(a-1)-1) \cdot
F_d\left(\frac{1}{\hat{r}\hat{s}}\right) -
(1-\hat{s}^2(b-1))(\hat{r}^2(a-1)-1)\cdot
F_d\left(\frac{\hat{s}}{\hat{r}}\right) \\
-\hat{s}^{4v+2} \hat{r}^2(b-1-\hat{s}^2)(a-1-\hat{r}^2)\cdot
F_d\left(\frac{\hat{r}}{\hat{s}}\right) +
\hat{r}^2(\hat{s}^2(b-1)-1)(a-1-\hat{r}^2) \cdot F_d(\hat{r}\hat{s}) \\
- \hat{s}^{{2v}}(\hat{s}^4-1)(\hat{r}^4-1).
\end{multline}

Up to this point, we have omitted the dependence of the parameters $a$ and $b$
in $F_d(x)$ for convenience. In full detail, $F_d(x) \equiv F_d(x;a,b)$. This
will be important in the next step when we look at the result of mapping $a
\leftrightarrow b$. For this, we define $G_d(x) = F_d(x;b,a)$.

With a little work we have 
\begin{align}
G_d(x) &= F_d(x;b,a)\\
&= x^{2v} F_d\left(\frac{1}{x};a,b\right).
\end{align}
Swapping $a \leftrightarrow b$ in equation~\Ref{eqn:abfuncteqn}, we get
\begin{multline}
0 = \hat{s}^{4v+2} (a-1-\hat{s}^2)(\hat{r}^2(b-1)-1) \cdot
G_d\left(\frac{1}{\hat{r}\hat{s}}\right) -
(1-\hat{s}^2(a-1))(\hat{r}^2(b-1)-1)\cdot
G_d\left(\frac{\hat{s}}{\hat{r}}\right) \\
-\hat{s}^{4v+2} \hat{r}^2(a-1-\hat{s}^2)(b-1-\hat{r}^2)\cdot
G_d\left(\frac{\hat{r}}{\hat{s}}\right) +
\hat{r}^2(\hat{s}^2(a-1)-1)(b-1-\hat{r}^2) \cdot G_d(\hat{r}\hat{s}) \\
- \hat{s}^{{2v}}(\hat{s}^4-1)(\hat{r}^4-1).
\end{multline}
Now convert $G_d$ back to $F_d$ using the relation $G_d(x) = 
x^{2v} F_d\left(\frac{1}{x}\right)$ and clear denominators to find
\begin{multline}
\label{eqn:newabfuncteqn}
0 = \hat{r}^{4v+2} (\hat{s}^2(a-1)-1)(b-1-\hat{r}^2) \cdot
F_d\left(\frac{1}{\hat{r}\hat{s}}\right) - \hat{r}^{4v+2}s^{2}
(a-1-\hat{s}^2)(b-1-\hat{r}^2)\cdot F_d\left(\frac{\hat{s}}{\hat{r}}\right) \\
-(\hat{s}^2(a-1)-1)(\hat{r}^2(b-1)-1)\cdot
F_d\left(\frac{\hat{r}}{\hat{s}}\right) +
\hat{s}^2(\hat{r}^2(b-1)-1)(a-1-\hat{s}^2) \cdot F_d(\hat{r}\hat{s}) \\
- \hat{r}^{{2v}}(\hat{s}^4-1)(\hat{r}^4-1).
\end{multline}

Combining equations~\Ref{eqn:abfuncteqn} and~\Ref{eqn:newabfuncteqn}, we can
eliminate one more boundary term 
(e.g. $F_d\left(\frac{1}{\hat{r}\hat{s}}\right)$) resulting in
\begin{align}
0 = \alpha(\hat{r}, \hat{s}) \cdot F_d\left(\frac{\hat{r}}{\hat{s}}\right) +
\beta(\hat{r}, \hat{s}) \cdot F_d\left(\frac{\hat{s}}{\hat{r}}\right) +
\gamma(\hat{r}, \hat{s}) \cdot F_d\left({\hat{r}}{\hat{s}}\right) +
\delta(\hat{r}, \hat{s}).
\end{align}
We do not state all of the coefficients $\alpha,\beta,\gamma$ (they are easily 
computed but complicated), however the coefficient $\delta$ will be important 
in what follows
\begin{multline}
 \delta =
r^{2v}s^{2v}(1-r^4)(1-s^4)
\big[ 
(1-b+r^2)(1+s^2-as^2)r^{2v+2}\\
-(1-b+s^2)(1+r^2-ar^2)s^{2v+2}
\big].
\label{eqn:abdelta}
\end{multline}
Note that for $a,b$ in this general case, $\delta = 0$ when $r,s$ are fourth 
roots of unity or $r=s$.

We follow the same logic as for the previous section. The locations of the
singularities are when the functions $\alpha = \beta = \gamma = 0$ and $\delta
\neq0$. Thus, solving for when $\alpha = \beta = \gamma = 0$ simultaneously
gives
\begin{align}
\label{eqn:rsabsoln}
r^{4v+4} 
&= \frac{(r^2(b-1) - 1)(r^2(a-1) - 1)}{(b-1-r^2)(a-1-r^2)} & 
\text{and} &&
s^{4v+4} &= \frac{(s^2(b-1) - 1)(s^2(a-1) - 1)}{(b-1-s^2)(a-1-s^2)}.
\end{align}
By rearranging equation~\Ref{eqn:rsabsoln}, we get
\begin{align}
0&= \hat{r}^{4v+4}\left({(b-1-\hat{r}^2)(a-1-\hat{r}^2)}\right) -
\left({(\hat{r}^2(b-1) - 1)(\hat{r}^2(a-1) - 1)}\right)\\
0&= \hat{s}^{4v+4}\left({(b-1-\hat{s}^2)(a-1-\hat{s}^2)}\right) -
\left({(\hat{s}^2(b-1) - 1)(\hat{s}^2(a-1) - 1)}\right)
\end{align}
which is in the form given in the Theorem \ref{thm:smalla} with $n=4$, $h(z) =
(b-1-z^2)(a-1-z^2)$ and $\lambda = 1$. The zeros of $h(z)$ are given by
\begin{equation}
z = \pm \sqrt{a-1}, \pm \sqrt{b-1}.
\end{equation}
Hence, the zeros of $h(z)$ will be inside the closed disc exactly when $a,b \leq
2$. Consequently when $a,b \leq 2$ we know that $\hat{r},\hat{s}$ lie on the 
unit circle. When $a$ or $b > 2$ we observe that all the solutions lie either 
on the unit circle or the real line.

\section{Exact and asymptotic results}
\begin{figure}[h!]
\begin{center}
	\includegraphics[width=0.60\textwidth]{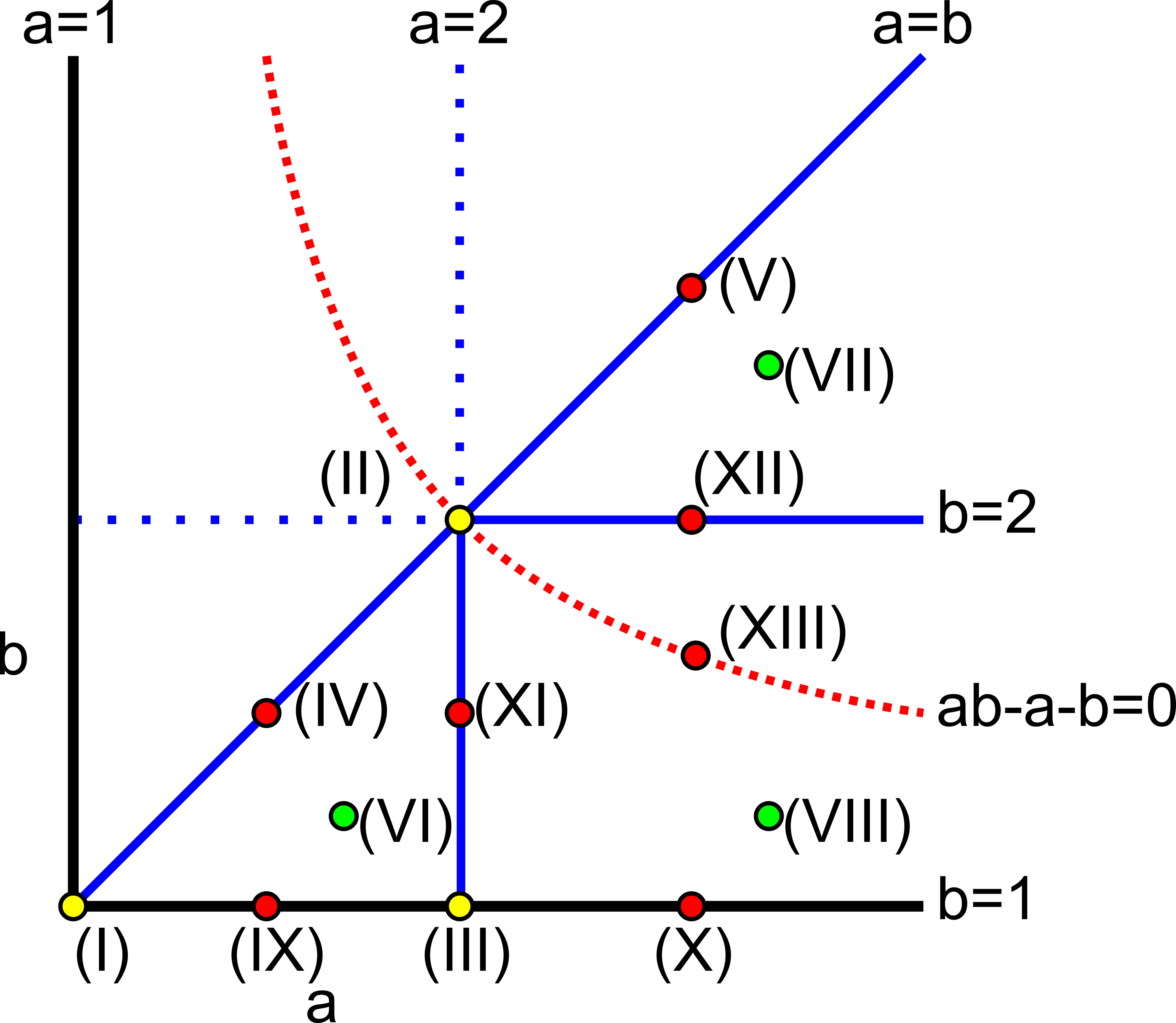}
\end{center}
\caption{The $a-b$ parameter space contains 13 representative points, depending
on whether $a,b=1$, $1<a,b<2$, $a,b=2$, $a,b>2$, or if $a=b$ or if $a,b$ lie on
along a special curve $ab=a+b$. The numbers in this diagram correspond to the
cases described in the text.}\label{fig param}
\end{figure}

In this section, we will describe the asymptotic and exact results we obtained
for each of the cases. In the case where both $a,b \in \{1,2\}$ or $ab=a+b$, we 
are able to obtain an exact solution for the dominant singularity. However, 
more generally we are only able to obtain asymptotic results. Note that by 
$a\leftrightarrow b$ symmetry, we need only consider cases where $a\geq b$. 
This gives 13 different cases (see Figure~\ref{fig param}) which we summarise 
in Section~\ref{subs:summary}.

In what follows we proceed by solving equation~\Ref{eqn:rsabsoln} for possible 
values of $\hat{r},\hat{s}$; we are able to do this exactly for a small number 
of cases, but in the majority we must do so asymptotically. Each pair of 
$\hat{r},\hat{s}$ may lead to a singularity of the generating function
however only when the auxiliary function $\delta$ is non-zero.

\subsection{Case (I) : $a = b =1$.}
\label{subs:case1}
\begin{mycase}\label{case1}\end{mycase}
This case is the non-interacting case. We can obtain the asymptotic expansion by
looking at equation~\Ref{eqn:a1b1soln} with $j=1,k=2$.
\begin{equation}
\label{eqn:a1b1asympt}
z_c = \frac{1}{4} + \frac{5}{32}\frac{\pi^2}{v^2} - \frac{5}{8}\frac{\pi^2}{v^3}
+ O\left(v^{-4}\right).
\end{equation}
		
\subsection{Case (II): $a = b = 2$.}
\label{subs:case2}
\begin{mycase}\label{case2}\end{mycase}

Simplifying the solutions for $\hat{r}$ and $\hat{s}$ in
equation~\Ref{eqn:rssoln}, we get that

\begin{equation}
\hat{r}^{4v} = \frac{1}{\hat{r}^4} \qquad \hat{s}^{4v} = \frac{1}{\hat{s}^4}.
\end{equation}
This suggests that the solutions for $\hat{r}$ and $\hat{s}$ are simple roots of
unity. Hence the set of solutions given by
\begin{align}
\hat{r} &\in \left\{\exp\left[ \frac{\pi i j}{2v + 2}\right]\right\}_{0\leq j 
\leq
4v+4} &
\hat{s} &\in \left\{\exp\left[ \frac{\pi i k}{2v + 2}\right]\right\}_{0\leq k 
\leq
4v+4}
\end{align}
is a superset of singularities for $\hat{r}$ and $\hat{s}$. 

If we attempt to set both $\hat{r}, \hat{s} = 1$, we do not obtain a valid 
singularity since $\delta = 0$. To obtain the dominant singularity, we 
instead take 
\begin{align}
\hat{r} &=\exp\left[ \frac{\pi i}{2v + 2}\right] &
\hat{s} &= 1,
\end{align}
and with this choice $\delta \neq 0$. Note that by symmetry we could also 
swap the choices of $\hat{r} \leftrightarrow \hat{s}$. We then have
\begin{align}
z_c &= \frac{1}{\left(\hat{r} + \frac{1}{\hat{r}}\right)\left(\hat{s} +
\frac{1}{\hat{s}}\right)} = \frac{1}{4 \cos\left(\frac{\pi}{2v+2}\right)}\nn
&= \frac{1}{4} + \frac{1}{32}\frac{\pi^2}{v^2} - \frac{1}{16}\frac{\pi^2}{v^3} +
O\left(v^{-4}\right).
\label{eqn:zc_case2}
\end{align}

\subsection{Case (III): $a = 2; \: b = 1$.}
\label{subs:case3}
\begin{mycase}\label{case3}\end{mycase}

As per the previous two cases, we find that the particular choice of $a$ and 
$b$ leads to solutions that are roots of unity. Equation~\Ref{eqn:rsabsoln} 
reduces to
\begin{equation}
\hat{r}^{4v} = -\frac{1}{\hat{r}^6} \qquad \hat{s}^{4v} = -\frac{1}{\hat{s}^6},
\end{equation}
and so the solutions are given by
\begin{align}
\hat{r} &= \left\{\exp\left[ \frac{\pi i j}{4v +
6}\right]\right\}_{\substack{0\leq j \leq 4v+4\\  {\rm j \; odd} }} &
\hat{s} &= \left\{\exp\left[ \frac{\pi i k}{4v +
6}\right]\right\}_{\substack{0\leq k \leq 4v+4\\  {\rm k \; odd} }}.
\end{align}

To obtain the dominant singularity we take $j,k=1,3$ respectively:
\begin{align}
\hat{r} &= \exp\left[ \frac{\pi i }{4v + 6}\right] &
\hat{s} &= \exp\left[ \frac{3\pi i}{4v + 6}\right],
\end{align}
and this gives a non-zero $\delta$
\begin{equation}
\delta = \frac{-6\pi^3}{v^3} + \frac{27\pi^3}{v^4} + \frac{\pi^3 (47\pi^2-1296)
}{16v^5} + O\left(v^{-6}\right).
\end{equation}
The dominant singularity is
\begin{equation}
z_c =
\frac{1}{4\cos\left(\frac{\pi}{4v+6}\right)\cos\left(\frac{3\pi}{4v+6}
\right) },
\end{equation}
and its asymptotic expansion is
\begin{equation}
\label{eqn:zc_case3}
z_c = \frac{1}{4} + \frac{5}{64}\frac{\pi^2}{v^2} -
\frac{15}{64}\frac{\pi^2}{v^3} + O\left(v^{-4}\right).
\end{equation}
Note that if we tried choosing $j,k=1,1$ then $\hat{r}=\hat{s}$ and $\delta = 
0$.

\subsection{Case (IV): $a = b; \: a < 2$.}
\label{subs:case4}
\begin{mycase}\label{case4}\end{mycase}

In Cases~\ref{case1} and~\ref{case2}, the solutions of $\hat{r}$ and
$\hat{s}$ are simply roots of unity. Hence we guess that for this generalised
case $1 < a=b < 2$, the solutions of $\hat{r}$ and $\hat{s}$ will be 
perturbations of the roots of unity found in the $a=b=1$ case (a similar 
approach was used in \cite{brak2005a-:a}). More precisely, we look for a 
solution of the form
\begin{align}
 \hat{r} &= \exp\left[ \frac{i \pi}{v+2} \left( c_0 + \frac{c_1}{v} + 
\frac{c_2}{v^2} +  \cdots \ \right) \right],
\end{align}
and similarly for $\hat{s}$. We substitute this into equation~\Ref{eqn:rssoln} 
and solve for the unknown constants. This process yielded
\begin{equation}
\hat{r} = \exp\left[\frac{i\pi}{v-\frac{2}{a-2}}\left(1 -
\frac{4a(a-1)\pi^2}{3(v(a-2)-1)^3} + O\left(\frac{1}{(
v(a-2)-2)^{5}}\right)\right)\right]
\end{equation}
which, when substituted into equation~\Ref{eqn:rssoln} gives
\begin{equation}
\hat{r}^{2v} - \frac{ \hat{r}^2 (a-1)-1 }{\hat{r}^2(a-1-\hat{r}^2)} =
\frac{8ia(a-1)(a^2+8a-8)\pi^5}{15(a-2)^4v^5} + O\left(v^{-6}\right).
\end{equation}
Repeating this for $\hat{s}$ leads to
\begin{equation}
\hat{s} = \exp\left[\frac{i\pi}{2\left(v-\frac{2}{a-2}\right)}\left(1 -
\frac{a(a-1)\pi^2}{3(-2+v(a-2))^3} + O\left(\frac{1}{(-2 +
v(a-2))^{5}}\right)\right)\right]
\end{equation}
which, when substituted into equation~\Ref{eqn:rssoln} gives
\begin{equation}
\hat{s}^{2v} + \frac{\hat{s}^2(a-1)-1 }{\hat{s}^2(a-1-\hat{s}^2)} =
\frac{ia(a-1)(a^2+8a-8)\pi^5}{60(a-2)^4v^5} + O\left(v^{-6}\right).
\end{equation}
Note that equation~\Ref{eqn:rssoln} is not symmetric under $\hat{r} 
\leftrightarrow \hat{s}$. 

This choice of $\hat{r}$ and $\hat{s}$ gives a $\delta$ value of
\begin{equation}
\delta = \hat{r}^{2v}\hat{s}^{2v}(\hat{r}^4-1)(\hat{s}^4-1) =
\frac{8\pi^2}{v^2}+\frac{8i\pi^2 (3a\pi-4i)}{(a-2)v^3} + O\left(v^{-4}\right)
\end{equation}
which is non-zero. Hence, using solving the kernel equation
$K(\hat{r},\hat{s})=0$ for $z$, we get that
\begin{equation}
\label{eqn:zcaasmall}
z_c =
\frac{1}{4}+\frac{5}{32}\frac{\pi^2}{v^2}+\frac{5}{8}\frac{\pi^2}{v^3(a-2)} +
O\left(v^{-4}\right).
\end{equation}
We see that as $a \to 1$ this agrees with Case~\ref{case1}.

\subsection{Case (V): $a = b; \: a  > 2$.}
\label{subs:case5}
\begin{mycase}\label{case5}\end{mycase}

In the case $a>2$, Theorem \ref{thm:smalla} does not hold and we expect
equation~\Ref{eqn:rssoln} to contain extra solutions along the real axis. By
rearranging equation~\Ref{eqn:rssoln}, we get that
\begin{align}
(a-1-\hat{r}^2)\hat{r}^{{2v}+2} &= \hat{r}^2 (a-1)-1,\\
(a-1-\hat{s}^2)\hat{s}^{{2v}+2} &= - \hat{s}^2(a-1)-1 .
\end{align}

We observe that $\hat{r} =\sqrt{a-1}$ will set the left hand side to zero and
leave a small remainder on the right. Hence, we looked at solutions that 
perturb this square root (again a similar approach was used in 
\cite{brak2005a-:a}). We proceed as per the previous case and arrive at a
solution of the form
\begin{align}
\hat{r} & = \sqrt{a-1} \left[1-\frac{a(a-2)}{2(a-1)^2 (a-1)^{v}} + O\left(v
(a-1)^{-2v} \right)\right]  \\
\hat{s} &= \frac{1}{\sqrt{a-1}} \left[1+\frac{a(a-2)}{2(a-1)^2 (a-1)^{v}} +
O\left(v (a-1)^{-2v} \right)\right].
\end{align}
This choice of $\hat{r}$ and $\hat{s}$ will give a non-zero $\delta$
which to leading order is
\begin{equation}
\delta = (a-1)^{2v} a^2 (a-2)^2 + O\left(v \right).
\end{equation}
Putting this together with the kernel equation $K(\hat{r},\hat{s})=0$ we get
\begin{equation}
z_c = \frac{a-1}{a^2} + \frac{(a-2)^2}{a^2(a-1) (a-1)^{v}}
+O\left(v(a-1)^{-2v}\right).
\end{equation}

\subsection{Case (VI): $a < 2; \: b < 2$.}
\label{sec:case6}
\begin{mycase}\label{case6}\end{mycase}

In Cases \ref{case1}, \ref{case2} and \ref{case4}, the
solutions of $\hat{r}$ and $\hat{s}$ are simple perturbations of roots of unity.
Hence we guess that for the case $1 < a,b < 2$, the solutions of $\hat{r}$ and
$\hat{s}$ will be of a similar nature. Hence we apply a similar method to that 
used in Case~\ref{case4} but now applied to equation~\Ref{eqn:rsabsoln}. 
This leads us to
\begin{multline}
\hat{r} = \exp\left[ 
\frac{\pi i}{\left(v- \frac{a+b-4}{(a-2)(b-2)}\right)}
\left(1 -
\frac{2(ab-a-b)(a^2b+ab^2-10ab+8a+8b-8)\pi^2}{3(a-2)^3(b-2)^3 \left(v-
\frac{a+b-4}{(a-2)(b-2)}\right)^3} \right.\right.\\
\left.\left.
%%%
\phantom{\frac{2(ab-a-b)(a^2b+ab^2-10ab+8a+8b-8)\pi^2}{3(a-2)^3(b-2)^3 \left(v-
\frac{a+b-4}{(a-2)(b-2)}\right)^3}}
%%%
+ O\left(\left(v-
\frac{a+b-4}{(a-2)(b-2)}\right)^{-5}\right) \right) \right];
\end{multline}
which, when substituted into equation~\Ref{eqn:rsabsoln} gives
\begin{equation}
\hat{r}^{4v+4} - \frac{(\hat{r}^2(b-1) - 1)(\hat{r}^2(a-1) -
1)}{(b-1-\hat{r}^2)(a-1-\hat{r}^2)} = O\left(\frac{1}{(a-2)^5 (b-2)^5
v^{5}}\right).
\end{equation}
We remind the reader that in this case if $\hat{r} =\hat{s}$ 
then $\delta = 0$ and so we need the value of $\hat{s}$ to be 
different. Following the same trend as for the previous case, we get that
\begin{multline}
\hat{s} = \exp\left[ \frac{\pi i}{2\left(v- \frac{a+b-4}{(a-2)(b-2)}\right)}
\left(1 - \frac{(ab-a-b)(a^2b+ab^2-10ab+8a+8b-8)\pi^2}{6(a-2)^3(b-2)^3 \left(v-
\frac{a+b-4}{(a-2)(b-2)}\right)^3} \right.\right.\\
%%%
\left.\left.
\phantom{
\frac{(ab-a-b)(a^2b+ab^2-10ab+8a+8b-8)\pi^2}{6(a-2)^3(b-2)^3 \left(v-
\frac{a+b-4}{(a-2)(b-2)}\right)^3}
}
%%% 
+ O\left(\left(v-
\frac{a+b-4}{(a-2)(b-2)}\right)^{-5}\right) \right)\right];
\end{multline}
which, when substituted into equation~\Ref{eqn:rsabsoln} gives
\begin{equation}
\hat{s}^{4v+4} - \frac{(\hat{s}^2(b-1) - 1)(\hat{s}^2(a-1) -
1)}{(b-1-\hat{s}^2)(a-1-\hat{s}^2)} = O\left(\frac{1}{(a-2)^5 (b-2)^5
v^{5}}\right).
\end{equation}
This choice of $\hat{r}$ and $\hat{s}$ will give a $\delta$ value of
\begin{equation}
\delta = -\frac{16\pi^2(a-2)(b-2)}{v^2} - \frac{16i \pi^2 (6ab\pi - 6b\pi - 2bi-
2ai -9a \pi +6\pi + 8i)}{v^3} + O\left(v^{-4}\right)
\end{equation}
which is non-zero. Hence, solving the kernel equation $K(\hat{r},\hat{s})=0$ for
$z$, we get that
\begin{equation}
\label{eqn:zcabsmall}
z_c = \frac{1}{4}+\frac{5}{32}\frac{\pi^2}{v^2}+\frac{5}{16}\frac{\pi^2 (a+b-4)
}{v^3(a-2)(b-2)} + O\left(v^{-4}\right).
\end{equation}

Note that equation~\Ref{eqn:zcabsmall} reduces to equation~\Ref{eqn:zcaasmall}
when $b=a$, and reduces to equation~\Ref{eqn:a1b1asympt} when $a,b \to 1$.

\subsection{Case (VII): $a > 2; \: b > 2$.}
\label{subs:case7}
\begin{mycase}\label{case7}\end{mycase}

In the case where $a$ or $b$ is greater than $2$, we argue as 
for Case~\ref{case5} in that we expect solutions along the real axis as 
well. Since $\hat{r}$ and $\hat{s}$ satisfy the same equation and the equation 
is invariant under switching $a$ and $b$, we can (without loss of generality) 
look at the expansion of $\hat{r}$ in terms of $\sqrt{a-1}$. We get
\begin{equation}
\label{eqn:largerabsoln}
\hat{r} = \sqrt{a-1} \left[1+ \frac{a(ab-a-b)(a-2)}{2(a-1)^3(a-b)(a-1)^{2v}}+
O\left( (a-1)^{-4v}\right) \right].
\end{equation}
Using the same process, we get that
\begin{equation}
\hat{s} = \sqrt{b-1} \left[1+ \frac{b(ab-a-b)(b-2)}{2(b-1)^3(b-a)(b-1)^{2v}}+
O\left( (b-1)^{-4v}\right) \right].
\end{equation}
We then check that this gives a non-zero value of $\delta$. For simplicity of 
notation, we let $A = a-1$ and $B= b-1$ and through abuse of notation, we 
obtain
\begin{equation}
\delta = A^{2v}B^{v} \left[A(AB-1)(A-B)(A^2-1)(B^2-1) + O\left(A^{-2v}\right) +
O\left(B^{-v}\right) \right].
\end{equation}
By making the substitution into $K(\hat{r},\hat{s})=0$, we get that to leading
order
\begin{multline}
z_c = \frac{\sqrt{a-1}\sqrt{b-1}}{ab} + \frac{(a-2)^2
(ab-a-b)\sqrt{b-1}}{2ab(b-a) \sqrt{a-1}(a-1)^{2v+2}} \\
+ \frac{(b-2)^2 (ab-a-b)\sqrt{a-1}}{2ab(a-b) \sqrt{b-1}(b-1)^{2v+2}}  +
O\left(a^{-4v}\right) + O\left(b^{-4v}\right).
\end{multline}
Note that the above expression implies that $z_c$ is a decreasing function of 
$v$. To see this, consider $a>b>2$. The first correction term is now negative 
(since $(b-a)<0$) while the second correction term is positive. The factor of 
$(a-1)^{2v+2}$ in the denominator of the first correction term is larger than 
the corresponding factor of $(b-1)^{2v+2}$ in the denominator of the second 
term. Hence for large $v$ the first correction term is smaller and negative 
than the larger and positive second correction term. Finally as $v\to \infty$ 
the sum of two corrections is positive and shrinking to $0$.

\subsection{Case (VIII): $a > 2; \: b < 2$.}
\label{subs:case8}
\begin{mycase}\label{case8}\end{mycase}
 
The next region we consider is when one parameter is small ($<2$) and the other 
is large ($>2$). Without loss of generality, we can assume that $a>2$
and $b<2$. We make use of the solutions obtained in Cases~\ref{case6} 
and~\ref{case7} to obtain
\begin{equation}
\hat{r} = \sqrt{a-1} \left[1+ \frac{a(ab-a-b)(a-2)}{2(a-1)^3(a-b)(a-1)^{2v}}+
O\left( (a-1)^{-4v}\right) \right]
\end{equation}
and
\begin{multline}
\hat{s} = \exp\left[ \frac{\pi i}{2\left(v- \frac{a+b-4}{(a-2)(b-2)}\right)}
\left(1 - \frac{(ab-a-b)(ab^2+a^2b-10ab+8a+8b-8)\pi^2}{6(a-2)^3(b-2)^3 \left(v-
\frac{a+b-4}{(a-2)(b-2)}\right)^3} \right.\right.\\
\left.\left.
%%%
\phantom{\frac{(ab-a-b)(ab^2+a^2b-10ab+8a+8b-8)\pi^2}{6(a-2)^3(b-2)^3 \left(v-
\frac{a+b-4}{(a-2)(b-2)}\right)^3}}
%%%
+ O\left(\left(v-
\frac{a+b-4}{(a-2)(b-2)}\right)^{-5}\right) \right)\right].
\end{multline}
Substituting these choices into $\delta$ give the following non-zero form
\begin{equation}
\delta = 2\pi A^{2v}(AB-1)(A^2-1) \left[ -\frac{i(A-1)}{v}
 +O(v^{-2})\right].
\end{equation}
We can then extract the growth rate as
\begin{equation}
z_c = \frac{\sqrt{a-1}}{2a} \left[1 + \frac{\pi^2}{8v^2} +
\frac{\pi^2(a+b-4)}{4(a-2)(b-2)v^3} + O\left(v^{-4}\right)\right].
\end{equation}
Note that as $b \to 1$ the above expression becomes 
equation~\Ref{eqn:zc_case10} in Case~\ref{case10} below.

We now complete the analysis by looking at the remaining boundary cases.
\subsection{Case (IX): $a < 2; \: b = 1$.}
\label{subs:case9}
\begin{mycase}\label{case9}\end{mycase}

In this case, equation~\Ref{eqn:rsabsoln} reduces down to
\begin{align}
\hat{r}^{4v+6} &= \frac{\hat{r}^2 (a-1) - 1 }{a-1-\hat{r}^2}  &
\hat{s}^{4v+6} &= \frac{\hat{s}^2 (a-1) - 1 }{a-1-\hat{s}^2}.
\end{align}

Following similar techniques used Section \ref{case6}, we can obtain the
two primitive roots of $\hat{r}$ and $\hat{s}$ to get
\begin{equation}
\hat{r} = \exp\left[\frac{\pi i}{2\left(v + \frac{a-3}{a-2} \right)}\left(1 -
\frac{a(a-1)\pi^2}{6\left(a-3 + v(a-2)\right)^3} + O\left(\frac{1}{(a-3 +
v(a-2))^5}\right)\right) \right]
\end{equation}
and
\begin{equation}
\hat{s} = \exp\left[\frac{\pi i}{\left(v + \frac{a-3}{a-2} \right)}\left(1 -
\frac{2a(a-1)\pi^2}{3\left(a-3 + v(a-2)\right)^3} + O\left(\frac{1}{(a-3 +
v(a-2))^5}\right)\right) \right].
\end{equation}
Using these values of $\hat{r}$ and $\hat{s}$, we obtain a non-zero $\delta$
value of
\begin{equation}
\delta = -\frac{16\pi^2 (a-2)}{v^2} - \frac{16 \pi^2i (2ia + 3\pi a - 6i)}{v^3}
+ O\left(v^{-4}\right).
\end{equation}
This will yield a dominant singularity of
\begin{equation}
z_c = \frac{1}{4} + \frac{5}{32}\frac{\pi^2}{v^2} - \frac{5}{16}
\frac{\pi^2(a-3)}{(a-2)v^3} + O\left(v^{-4}\right).
\end{equation}
As $a \to 1$ this reduces to equation~\Ref{eqn:a1b1asympt}.

\subsection{Case (X): $a > 2; \: b = 1$.}
\label{subs:case10}
\begin{mycase}\label{case10}\end{mycase}

In this case, we have the same equations for $\hat{r}$ and $\hat{s}$ as the 
previous case
\begin{align}
\hat{r}^{4v+6} &= \frac{\hat{r}^2 (a-1) - 1 }{a-1-\hat{r}^2} &
\hat{s}^{4v+6} &= \frac{\hat{s}^2 (a-1) - 1 }{a-1-\hat{s}^2}.
\end{align}
Following methods used in Cases~\ref{case5} and~\ref{case7},
we can obtain the singularity of $\hat{r}$ on the real line:
\begin{equation}
\hat{r} = \sqrt{a-1} \left[1 - \frac{a(a-2)}{2(a-1)^4 (a-1)^v}
+O\left(v(a-1)^{-2v}\right) \right]
\end{equation}
while
\begin{equation}
\hat{s} = \exp\left[\frac{\pi i}{2\left(v + \frac{a-3}{a-2} \right)}\left(1 -
\frac{a(a-1)\pi^2}{6\left(a-3 + v(a-2)\right)^3} + O\left(\frac{1}{\left(a-3 +
v(a-2)\right)^5}\right)\right) \right].
\end{equation}

Using these values of $\hat{r}$ and $\hat{s}$, we obtain a non-zero $\delta$
\begin{equation}
\delta = \frac{2i a\pi (a-1)^{2v+2}(a-2)^2}{v}
+ O\left(\frac{(a-1)^{2v}}{v^{2}}\right).
\end{equation}

This yields a dominant singularity as
\begin{equation}
\label{eqn:zc_case10}
z_c = \frac{\sqrt{a-1}}{2a} \left[1 + \frac{\pi^2}{8v^2} -
\frac{(a-3)\pi^2}{4(a-2)v^3} + O\left(v^{-4}\right)\right].
\end{equation}

\subsection{Case (XI): $a = 2; \: b < 2$.}
\label{subs:case11}
\begin{mycase}\label{case11}\end{mycase}
This case is very similar to that of Case \ref{case9}.
equation~\Ref{eqn:rsabsoln} reduces down to
\begin{align}
\hat{r}^{4v+4} &= -\frac{\hat{r}^2 (b-1) - 1 }{b-1-\hat{r}^2} &
\hat{s}^{4v+4} &= -\frac{\hat{s}^2 (b-1) - 1 }{b-1-\hat{s}^2}.
\end{align}
Again we follow the method used in Case~\ref{case6}, and we find
\begin{align}
\hat{r} &= \exp\left[\frac{\pi i}{4\left(v + \frac{b-4}{b-2} \right)}\left(1 -
\frac{b(b-1)\pi^2}{3\left(b-4 + 2v(b-2)\right)^3}+ O\left(\frac{1}{\left(b-4 +
2v(b-2)\right)^5}\right)\right) \right]
\\
\hat{s} &= \exp\left[\frac{3 \pi i}{4\left(v + \frac{b-4}{b-2} \right)}\left(1 -
\frac{3b(b-1)\pi^2}{\left(b-4 + 2v(b-2)\right)^3} + O\left(\frac{1}{\left(b-4 +
2v(b-2)\right)^5}\right)\right) \right].
\end{align}
These give a non-zero $\delta$:
\begin{equation}
\delta = \frac{6\pi^3 (b-2)}{v^3} - \frac{3i \pi^3 (3ib + 6\pi b - 12i -
4\pi)}{v^4} + O\left(v^{-5}\right).
\end{equation}
And so we find the dominant singularity:
\begin{equation}
z_c = \frac{1}{4} + \frac{5}{64}\frac{\pi^2}{v^2} - \frac{5}{64}
\frac{\pi^2(b-4)}{(b-2)v^3} + O\left(v^{-4}\right).
\end{equation}		
Note that as $b \to 1$ we recover equation~\Ref{eqn:zc_case3}.

\subsection{Case (XII): $a > 2; \: b = 2$.}
\label{subs:case12}
\begin{mycase}\label{case12}\end{mycase}
As per Case \ref{case11}, we assume that $b=2$. This
reduces equation~\Ref{eqn:rsabsoln}
\begin{equation}
\hat{s}^{4v+4} = -\frac{\hat{s}^2(a-1)-1}{a-1-\hat{s}^2}.
\end{equation}
Looking at the expansion of $\hat{s}$, we get
\begin{multline}
\hat{s} = \exp\left[   \frac{\pi i}{2\left(2v - \frac{a-4}{a-2}\right)} 
\left(1- \frac{\pi^2(a-1)a}{3( (2a-4)v + a-4)^3} \right.\right.\\
\left.\left. 
%%%
\phantom{\frac{\pi^2(a-1)a}{3( (2a-4)v + a-4)^3}}
%%%
+ O\left(\frac{1}{\left( (2a-4)v +a-4 \right)^5}\right)\right)\right]
\end{multline}
Similarly, the solution for $\hat{r}$ is given by a simplified version of
equation~\Ref{eqn:largerabsoln}.
\begin{equation}
\hat{r} = \sqrt{a-1}\left[1 + \frac{a(a-2)}{2(a-1)^3(a-1)^{2v}} +
O\left((a-1)^{-4v}\right)\right].
\end{equation}
Together these give
\begin{equation}
\delta = (a-1)^{2v} \left[ \frac{\pi a (a-1)(a-2)^3}{v} 
+ O\left(v^{-2}\right)\right]
\end{equation}
with the dominant singularity being
\begin{equation}
z_c = \frac{\sqrt{a-1}}{2a} \left[ 1 +\frac{\pi^2}{32 v^2} -
\frac{(a-4)\pi^2}{32(a-2)v^3} + O\left(v^{-4}\right)\right].
\end{equation}

\subsection{Case (XIII): $ ab-a-b = 0$.}
\label{subs:case13}
\begin{mycase}\label{case13}\end{mycase}
Looking at Cases \ref{case6}, \ref{case7} and \ref{case8}, the
factor $ab-a-b$ appears in the asymptotic expansions, leading us to believe that
there may be something of interest along this line. We note that this 
polynomial plays an important role in the single-walk version of this model 
\cite{brak2005a-:a} --- along the curve $ab=a+b$ the dominant singularity is 
independent of the width of the system. While  this is not the case for the 
two-walk model we consider in this paper, we are able to compute the dominant 
singularity exactly along the curve.

equation~\Ref{eqn:rsabsoln} reduces down to
\begin{align}
(\hat{r}^2(a-1) - 1)(a-1-\hat{r}^2)(\hat{r}^{2v+2}-1)(\hat{r}^{2v+2}+1) &= 0,\\
(\hat{s}^2(a-1) - 1)(a-1-\hat{s}^2)(\hat{s}^{2v+2}-1)(\hat{s}^{2v+2}+1) &= 0.
\end{align}

This suggests that the solutions of $\hat{r}$ or $\hat{s}$ come in two forms.
One is a simple root of unity and the other is a square root type singularity. 
Again, the condition $\delta \neq 0$ requires $\hat{r} \neq \hat{s}$ and we 
obtain the following exact expressions
\begin{align}
\hat{r} &= \sqrt{a-1}\\
\hat{s} &= \exp\left(\frac{\pi i}{2v + 2}\right).
\end{align}
We could equally well have chosen the above with $\hat{r}$ and $\hat{s}$ 
swapped. Using the above values of $\hat{r}$ and $\hat{s}$, we obtain
\begin{equation}
\delta = (a-1)^{2v} \left[\frac{-2i a^2(a-2)^3 \pi}{v} + \frac{2(-2i+ia-\pi +\pi
a)\pi(a-2)^2a^2}{v^2} + O\left(v^{-3}\right)\right].
\end{equation}
This will yield a dominant singularity of
\begin{equation}
z_c = \frac{\sqrt{a-1}}{2a \cos\left(\frac{\pi}{2v+2}\right)}
\end{equation}
or asymptotically,
\begin{equation}
z_c = \frac{\sqrt{a-1}}{2a}\left[1 + \frac{\pi^2}{8v^2} - \frac{\pi^2}{4v^3}   +
O\left(v^{-4}\right) \right].
\end{equation}
Note that as $a \to 2$ this reduces to equation~\Ref{eqn:zc_case2}.

\subsection{Summary}
\label{subs:summary}
\renewcommand{\arraystretch}{1.3}
Here we simply summarise the results of this section and divided them into 
three tables. In Table~\ref{tab:zcexact} we give the cases in which we are able 
to find the dominant singularity exactly. For the remainder of the parameter 
space we have been unable to find exact expressions and we present only 
asymptotic results. These are divided into Tables~\ref{tab:zcsmall} 
and~\ref{tab:zcbig} according to whether or not at least one $a,b$ exceeds $2$. For comparison we include the asymptotics of the single-walk model with $b=1$ in Table~\ref{tab:zcsingle}.

\begin{table}[h!]
\begin{center}
\begin{tabular}{|| c | c | c | l||}
\hline
\hline
Case: & $a$ & $b$ & Dominant Singularity ($z_c$)\\
\hline
\hline
 (I) & $=1$ & $=1$ &
$=\frac{1}{4\cos\left(\frac{\pi}{2v+4}\right)\cos\left(\frac{2
\pi}{2v+4}\right)}$\\
 & & &$=\frac{1}{4} + \frac{5}{32}\frac{\pi^2}{v^2} -
\frac{5}{8}\frac{\pi^2}{v^3} + O\left(v^{-4}\right)$\\
 \hline
 (II) & $=2$ & $=2$ & $=\frac{1}{4\cos\left(\frac{\pi}{2v+2}\right)}$\\
 & & &$=\frac{1}{4} + \frac{1}{32}\frac{\pi^2}{v^2} -
\frac{1}{16}\frac{\pi^2}{v^3} + O\left(v^{-4}\right)$\\
\hline
 (III) & $=2$ & $=1$ &
$=\frac{1}{4\cos\left(\frac{\pi}{4v+6}\right)\cos\left(\frac{3\pi}{4v+6}\right)}
$\\
 & & &$=\frac{1}{4} + \frac{5}{64}\frac{\pi^2}{v^2} -
\frac{15}{64}\frac{\pi^2}{v^3} + O\left(v^{-4}\right)$\\
 \hline
 (XIII) & \multicolumn{2}{|c|}{$ab=a+b$} &
$=\frac{\sqrt{a-1}}{2a\cos\left(\frac{\pi}{2v+2}\right)}$\\
& \multicolumn{2}{|c|}{} &
$=\frac{\sqrt{a-1}}{2a}\left(1 + \frac{\pi^2}{8v^2} - \frac{\pi^2}{4v^3}  
+ O\left(v^{-4}\right) \right)$\\
 \hline
 \hline
\end{tabular}
\end{center}
\caption{The exact value and asymptotic behaviour of the dominant singularity 
when $a,b \in {1,2}$ and $ab=a+b$. Note that in each case $z_c$ decreases with 
increasing $v$. }
\label{tab:zcexact}
\end{table}

\begin{table}[h!]
\begin{center}
\begin{tabular}{|| c | c | c | l ||}
\hline
\hline
Case: & $a$ & $b$ & Dominant Singularity ($z_c$)\\
\hline
\hline
 (IV) &\multicolumn{2}{|c|}{$a=b<2$} & $=\frac{1}{4} + 
\frac{5}{32}\frac{\pi^2}{v^2} +
\frac{5}{8}\frac{\pi^2}{v^3(a-2)} + O\left(v^{-4}\right)$\\
 \hline
 (VI) & $<2$ & $<2$ & $=\frac{1}{4} + \frac{5}{32}\frac{\pi^2}{v^2} +
\frac{5}{16}\frac{\pi^2(a+b-4)}{v^3(a-2)(b-2)} + O\left(v^{-4}\right)$\\
\hline
 (IX) & $<2$ & $=1$ & $=\frac{1}{4} + \frac{5}{32}\frac{\pi^2}{v^2} -
\frac{5}{16}\frac{\pi^2(a-3)}{v^3(a-2)} + O\left(v^{-4}\right)$\\
 \hline
 (XI) & $=2$ & $<2$ & $=\frac{1}{4} + \frac{5}{64}\frac{\pi^2}{v^2} -
\frac{5}{64}\frac{\pi^2(b-4)}{v^3(b-2)} + O\left(v^{-4}\right)$\\
 \hline
 \hline
\end{tabular}
\end{center}
\caption{The asymptotic behaviour of the dominant singularity when $a,b \leq 
2$. Again note that in each case, $z_c$ is a decreasing function of $v$ and 
that $z_c \to \frac{1}{4}$ as $v \to \infty$.}
\label{tab:zcsmall}
\end{table}

\begin{table}[h!]
\begin{center}
\begin{tabular}{|| c | c | c | l||}
\hline
\hline
Case: & $a$ & $b$ & Dominant Singularity ($z_c$)\\
\hline
\hline
 (V) & \multicolumn{2}{|c|}{$a=b>2$} & $=\frac{a-1}{a^2} + 
\frac{(a-2)^2}{a^2(a-1)(a-1)^v} +
O\left(v(a-1)^{2v}\right)$\\
 \hline
 (VII) & $>2$ & $>2$ & $=\frac{\sqrt{a-1}\sqrt{b-1}}{ab}$ \\
& & & $+ \frac{(a-2)^2 (ab-a-b)\sqrt{b-1}}{2ab(b-a) \sqrt{a-1}(a-1)^{2v+2}}
 + \frac{(b-2)^2 (ab-a-b)\sqrt{a-1}}{2ab(a-b) \sqrt{b-1}(b-1)^{2v+2}} $ \\
& & & $+ O\left(a^{-4v}\right) + O\left(b^{-4v}\right)$\\
\hline
 (VIII) & $>2$ & $<2$ & $= \frac{\sqrt{a-1}}{2a} \left[1 +
\frac{1}{8}\frac{\pi^2}{v^2}+\frac{1}{4}\frac{\pi^2(a+b-4)}{(a-2)(b-2)v^3}
+O\left(v^{-4}\right) \right]$\\
 \hline
 (X) & $>2$ & $=1$ & $= \frac{\sqrt{a-1}}{2a} \left[1 +
\frac{1}{8}\frac{\pi^2}{v^2}-\frac{1}{4}\frac{\pi^2(a-3)}{(a-2)v^3}
+O\left(v^{-4}\right) \right]$\\
 \hline
 (XII) & $>2$ & $=2$ & $= \frac{\sqrt{a-1}}{2a} \left[1 +
\frac{1}{32}\frac{\pi^2}{v^2}-\frac{1}{32}\frac{\pi^2(a-4)}{(a-2)v^3}
+O\left(v^{-4}\right) \right]$\\
 \hline
 \hline
\end{tabular}
\end{center}
\caption{The asymptotic behaviour of the dominant singularity when at least one 
of $a,b > 2$. Note that $z_c$ decreases with increasing $v$ in all cases.}
\label{tab:zcbig}
\end{table}

\begin{table}
 \begin{center}
  \begin{tabular}{||c|c|l||}
\hline
\hline
   $a$ & $z_c$ & Asymptotic expansion \\
\hline\hline
  $1$ & $\frac{1}{2\cos\left(\frac{\pi}{2v+2} \right)} $ & 
$\sim \frac{1}{2}+\frac{\pi^2}{16v^2}-\frac{\pi^2}{8v^3} + 
O\left(v^{-4}\right)$\\
\hline
  $(1,2)$ & $\circ$ & $\sim 
\frac{1}{2}+\frac{\pi^2}{16v^2} - \frac{\pi^2}{8(2-a)v^3}+ 
O\left(v^{-4}\right)$\\
\hline
  $2$ & $ \frac{1}{2 \cos\left( \frac{\pi}{4v+2} \right)}$ & 
$ \sim 
\frac{1}{2} + \frac{\pi^2}{64v^2}-\frac{\pi^2}{64v^3} + O\left(v^{-4}\right)$ \\
\hline
  $(2,\infty)$ & $\circ$ & $\sim \frac{\sqrt{a-1}}{a}\left(
1 + \frac{(a-2)^2}{2(a-1)^{2v+2}} \right) + O\left( (a-1)^{-4v} \right) $  \\
\hline
\hline
  \end{tabular}
 \end{center}
\caption{The dominant singularity when $b=1$ for the single-walk model.}
\label{tab:zcsingle}
\end{table}

\section{Overview and discussion}
\subsection{Infinite slit phase diagram}

Recall that in the single walk case, discussed in the introduction, the order 
of the limits polymer length $n$ and slit width $w$ going to infinity matters; 
it was shown in \cite{brak2005a-:a} that 
 \begin{equation}
 \kappa^{single}_{half-plane}(a) \neq \kappa^{single}_{inf-slit}(a,b).
 \end{equation}
In fact the phase diagram for the single walk in the infinite slit, given in 
Figure~\ref{phase-force-diagram-single}(left), depends on both $a$ and $b$ 
whereas the half plane limit depends only on $a$. This can be understood by 
observing that a finite Dyck path must visit the bottom wall as it is fixed at 
both ends there so once the width is sent to infinity any finite Dyck path only 
feels the bottom wall, while if the length of the Dyck path is first sent to 
infinity the walk will ``see'' both walls for any finite width.

\begin{figure}[ht!]
\begin{center}
\includegraphics[width=8cm]{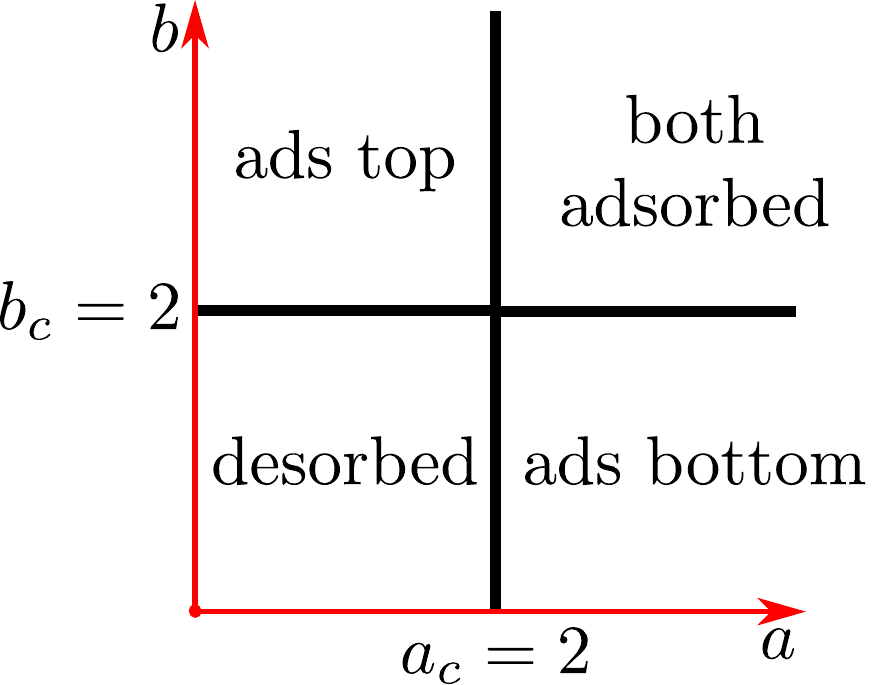}
\caption{
Phase diagram of the infinite strip for the two walk model analysed in this 
paper. There are four phases: a desorbed phase, a phase where the bottom walk is 
adsorbed
onto the bottom wall,  a phase where the top walk is adsorbed
onto the top wall, and a phase where both walks are adsorbed onto their 
respective walls.}
\label{phase-diagram} 
\end{center}
\end{figure}

From the calculations in the previous section we see that for the two walk 
model the infinite slit free energy is
\begin{align}
        \kappa_{inf-slit}(a,b) &=
\begin{cases}
 \log\left(4\right) & \mbox{ if } a,b \leq 2 \\ 
\log\left(\frac{2a}{\sqrt{a-1}}\right) & \mbox{ if } a > 2 \mbox{ and }
        b< 2 \\ 
\log\left(\frac{2b}{\sqrt{b-1}}\right) & \mbox{ if } a < 2 \mbox{ and }
        b >  2\\ 
\log\left(\frac{ab}{\sqrt{a-1}\sqrt{b-1}}\right) &  \mbox{ if } a \geq 2 \mbox{ 
and } b \geq  2 .
\end{cases}
\label{inf-slit-free-energy}
\end{align}
Hence the phase diagram can be illustrated as in Figure~\ref{phase-diagram}.

We observe that
\begin{equation}
     \kappa_{inf-slit}(a,b) =  \kappa^{single}_{half-plane}(a) 
+\kappa^{single}_{half-plane}(b)
\label{inf-strip-free-energy-split}
\end{equation}
and recalling equation~\Ref{double-half-plane-free-energy} we see that 
\begin{equation}
     \kappa_{inf-slit}(a,b) =   \kappa_{double-half-plane}(a,b) .
\label{inf-strip-half-plane-equality}
\end{equation}
So the free energy for this two walk model does \emph{not} depend on the 
order of the limits! 

This conclusion depends on the particular model we have chosen where both walks 
start on different walls. Had we considered a model where both walks 
started on the bottom wall this observation would be different; by taking the 
width to infinity first, neither walk would interact with the top wall and the 
free energy of this system would be that of two walks in a single half-plane. 
On the other hand, the infinite slit free energy does not depend on the end 
points of the polymer because the length is taken to infinity first.

\subsection{Force between the walls}

Using the asymptotic expressions for $\kappa$ found above  we obtain the 
asymptotics for the force. We have
  \begin{itemize}
  \item For $a,b<2$
    \begin{equation}
\mathcal{F}\sim\frac{5\pi^2}{w^3} ;
    \end{equation}
 \item For $a<2, b=2$
 \begin{equation}
\mathcal{F}\sim\frac{5\pi^2}{2 w^3} ;
    \end{equation}
  \item For $a<2,b>2$
    \begin{equation}
\mathcal{F}\sim\frac{\pi^2}{w^3} ;
    \end{equation}
     \item For $a>2,b<2$
    \begin{equation}
\mathcal{F}\sim\frac{\pi^2}{w^3};
    \end{equation}
 \item For $a=2, b<2$
    \begin{equation}
\mathcal{F}\sim\frac{5\pi^2}{2 w^3} ;
    \end{equation}
  \item For $a=2, b=2$
    \begin{equation}
\mathcal{F}\sim\frac{\pi^2}{w^3} ;
    \end{equation}
 \item For $a>2, b=2$
    \begin{equation}
\mathcal{F}\sim \frac{\pi^2}{4 w^3} ;
    \end{equation}
 \item For $a=2, b>2$
    \begin{equation}
\mathcal{F}\sim\frac{\pi^2}{4 w^3} ;
    \end{equation}
 \item For $a,b>2$ with $a>b$
    \begin{equation}
\mathcal{F} \sim \frac{(b-2)^2(ab-a-b) \log (b-1)}{2 (a-b)(b-1)^3} 
\left(\frac{1}{b-1}\right)^{w};
      \end{equation}
     \item For $a,b>2$ with $a<b$
    \begin{equation}
    \mathcal{F} \sim \frac{(a-2)^2(ab-a-b) \log (a-1)}{2 (b-a)(a-1)^3} 
\left(\frac{1}{a-1}\right)^{w};
      \end{equation}
  \item For $b=a>2$
    \begin{equation}
      \mathcal{F} \sim \frac{(a-2)^2 \log(a-1) }{2(a-1)^2} 
\left(\frac{1}{a-1}\right)^{w/2}.
    \end{equation}
  \end{itemize}
For any $a,b$ the force is positive and so is repulsive. This is in contrast to 
the single walk case where there is a region of attractive forces. The regions 
of the plane which gave different asymptotic expressions
for $\kappa$ and hence different phases for the infinite slit clearly
also give different force behaviours. There is also a special subtle change of 
the magnitude of the force when $a=b$ for $a,b>2$. On the other hand the special 
super-integrable curve $a+b=ab$ does not display special behaviour, which relates to which walk is less bound to its respective surface and 
so drives the value of the force, except when 
$a=b=2$. 

\begin{figure}[ht!]
\begin{center}
\includegraphics[width=10cm]{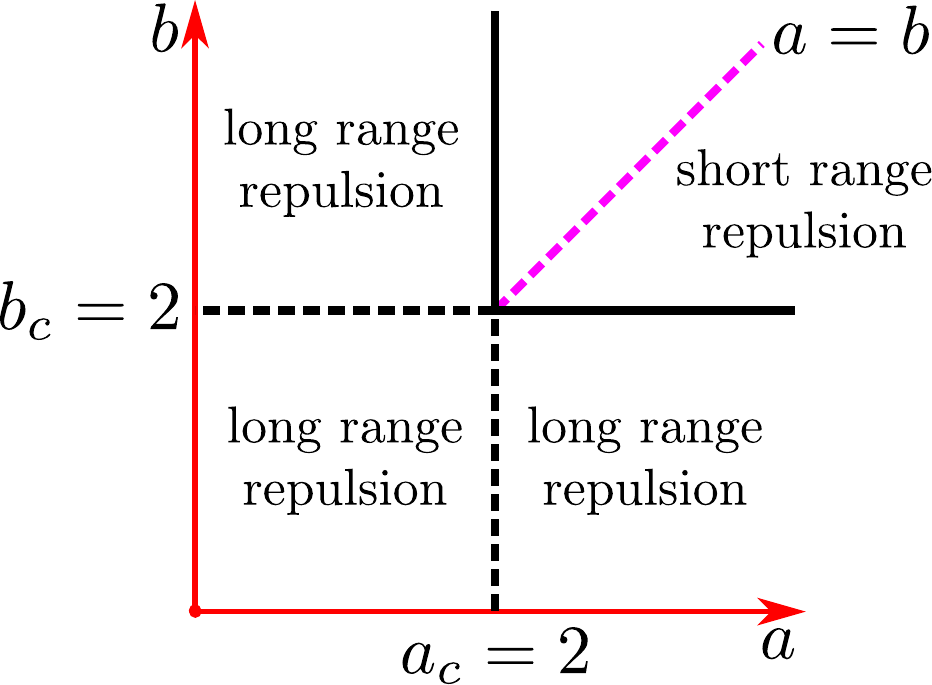}
\caption{ A diagram of the regions of different types of effective 
force between the walls of a slit. Short range behaviour refers to
exponential decay of the force with slit width while long range refers 
to a power law decay. On full lines there is a change from long to short range 
force decay. On the dashed lines there is a singular change of behaviour of the 
magnitude of the force. }
\label{force-diagram} 
\end{center}
\end{figure}

The difference between the single and two walk models can be understood as 
follows. When there are two walks they effective shield each other from the 
interactions of the other wall and it is when a single walk is sufficiently attracted to the 
two sides of the slit simultaneously that an attractive force eventuates. There 
are however changes in the magnitude and the range of the repulsive force arising 
from whether the walks are adsorbed or desorbed. When either or both walks are 
desorbed there is a long range force arising from the entropy of the walk(s) 
while if both are adsorbed the force is short-ranged as the excursions of either 
walk from the walls are relatively short-ranged.
The force diagram is given in Figure~\ref{force-diagram}.

\subsection{Conclusion}

A model of two polymers confined to be in a long macroscopic sized slit with 
sticky walls has been modelled by a directed walk system. Our results show 
distinct differences from the earlier single polymer results. In particular, we see differences from the single polymer system in both the phase diagram, and the sign and strength of the entropic force exerted by the polymers on the walls of the slit. 

The phase diagram contains four phases, whereas that the single walk model has only three. Moreover, this phase diagram is independent on the order one considers the limits of large width and length to be taken. This is also in contrast with the single walk system.

The force induced by the polymers remains repulsive in all parts of 
the phase diagram even though the range of the force does depend on whether the 
walks are adsorbed onto the walls. This again is in contrast with the single polymer system where an attractive regime is observed.  In our two polymer system each polymer is effectively shielded from the opposite wall by the other polymer. This gives rise to the difference between the results seen here and those of the single polymer system.

While we have a model that goes beyond the single polymer results, to obtain a situation which might replicate the non-directed self-avoiding polygon results of Alvarez
{\it et al.\ }\cite{alvarez2008self} one will need to allow both walks to 
interact with both walls. This will be significantly more complicated combinatorially to analyse.  

\section*{Acknowledgements}
Financial support from the Australian Research Council via its support for the 
Centre of Excellence for Mathematics and Statistics of Complex Systems and 
through its Discovery Program is gratefully acknowledged by the authors. 
Financial support from the Natural Sciences and Engineering Research Council of 
Canada (NSERC) though its Discovery Program is gratefully acknowledged by the 
authors. ALO thanks the University of British Columbia, and in particular Prof Mark Mac Lean, for hospitality.

\appendix
\section{Appendix: Proof of Theorem 1}

\begin{proof}
Let $\hat{r} = x e^{i t}$ for some $x >0$ and $0\leq t < 2\pi$. Substituting
this into the expression and manipulating gives
\begin{align}
\frac{(x e^{i t})^{v+2} - \frac{1}{(x e^{i t})^{v+2}}} {(x e^{i t})^v -
\frac{1}{(x e^{i t})^v}}
&= \frac{ (x^{v+2} - x^{-(v+2)}) \cos((v+2)t) + i (x^{v+2} + x^{-(v+2)})
\sin((v+2)t) } { (x^{v} - x^{-v})\cos(vt) + i (x^{v} + x^{-v}) \sin(vt)}.
\end{align}
By multiplying the denominator by its complex conjugate we obtain an expression
of the form $\left( P(x)+iQ(x) \right)/ D(x)$ and
\begin{align}
P(x)	&=  (x^{2v+2} + x^{-(2v+2)})\cos(2t) - (x^{2} + x^{-2})\cos((2v+2)t)\\
Q(x)	&=  (x^{2v+2} - x^{-(2v+2)})\sin(2t) - (x^{2}- x^{-2})(\sin((2v+2)t)\\
D(x)  &= (x^{2v} + x^{-2v}) - 2 \cos(2vt).
\end{align}
Note that $P,Q,D$ are all real. Hence this expression is real if and only if 
$Q(x) =0$. It is clear that if $\hat{r} \in \mathbb{R}$ ($t = 0,\pi$) or if
$\hat{r}$ is a complex number of unit magnitude ($x=1$), then $Q(x) =0$. Thus,
suppose that there is a value $\hat{r}$ that does not satisfy either case ($t
\neq 0, \pi$ and $x \neq 1$), then
\begin{equation}
0 = Q(x) =  (x^{2v+2} - x^{-(2v+2)})\sin(2t) - (x^{2}- x^{-2})(\sin((2v+2)t),
\end{equation}
which gives
\begin{equation}
\frac{x^{2v+2} - x^{-(2v+2)}}{x^{2}- x^{-2}} = \frac{\sin((2v+2)t)}{\sin(2t)}.
\end{equation}
The left hand side can be expanded to give the sum
\begin{equation}
\frac{x^{2v+2} - x^{-(2v+2)}}{x^{2}- x^{-2}} = x^{-2v} \sum_{i=0}^{v} x^{4i}
\end{equation}
with $v+1$ summands. When $v$ is even, the sum expands to
\begin{equation}
x^{2v} + x^{2v-4 } + \ldots + x^{4} + 1 + x^{-4} + \ldots + x^{-2v + 4} +
x^{-2v}
\end{equation}
and in the case where $v$ is odd, the sum expands to
\begin{equation}
x^{2v} + x^{2v-4 } + \ldots + x^{6} + x^{2} + x^{-2} + x^{-6} + \ldots + x^{-2v
+ 4} + x^{-2v}.
\end{equation}
In each case, the summands can be pairs off in the form $x^{2l} + x^{-2l}$ for
the appropriate values of $l$ and a remaining $1$ when $v$ is even. Now, for for
$x \neq 1$ and a positive integer $k$, we have $x^{k} + x^{-k} > 2$. Summing
over all pairs, we get
\begin{equation}
\frac{x^{2v+2} - x^{-(2v+2)}}{x^{2}- x^{-2}} > v+1.
\end{equation}

For the right hand side, we have
\begin{align}
\frac{\sin((2v+2)t)}{\sin(2t)} &= \frac{e^{i(2v+2)t} - e^{-i(2v+2)t}}{e^{i2t} -
e^{-i2t}},\\
\intertext{and by substituting $q= e^{2it}$ we get}
\frac{\sin((2v+2)t)}{\sin(2t)} &= \frac{q^{v+1} - q^{-(v+1)}}{q - q^{-1}} .
\end{align}
When expanded, this gives
\begin{equation}
\frac{q^{v+1} - q^{-(v+1)}}{q - q^{-1}} = q^{-v} \sum_{i=0}^{v} q^{2i}.
\end{equation}
Similar to the case with $x$, When $v$ is even, this sum expands to
\begin{equation}
q^{v} + q^{v-2} + \ldots + q^{2} + 1 + q^{-2} + \ldots + q^{-v+2} + q^{-v}
\end{equation}
and in the case where $v$ is odd, the sum expands to
\begin{equation}
q^{v} + q^{v-2} + \ldots + q^{3} + q + q^{-1} + q^{-3} + \ldots + q^{-v+2} +
q^{-v}.
\end{equation}
In either case, the powers of $q$ can be paired up and simplified as follows
\begin{equation}
q^{l} + q^{-l} = e^{2ilt} + e^{-2ilt} = 2\cos(2lt).
\end{equation}
Thus
\begin{equation}
\displaystyle
\frac{\sin((2v+2)t)}{\sin(2t)} = \left\{
\begin{array}{lr}
	 \ds 1+ 2 \sum_{j=1}^{v/2} \cos(2jt) & v \;{\rm even} \\
	\ds 2 \sum_{j=1}^{(v-1)/2} \cos(2(2j+1)t) & v \;{\rm odd}.
\end{array}
\right.
\end{equation}
In either case, each summand can be bounded above by $1$ and given the number of
summands, we can conclude that
\begin{equation}
\frac{x^{2v+2} - x^{-(2v+2)}}{x^{2}- x^{-2}} > v+1 \geq
\frac{\sin((2v+2)t)}{\sin(2t)}.
\end{equation}
Thus contradicting the existence of the point $\hat{r} = x e^{it}$ with $x\neq
1$ and $t \neq 0,\pi$.
\end{proof}

\bibliography{project}
\bibliographystyle{aip}

\end{document}